\documentclass[journal]{vgtc}               % review (journal style)
\onlineid{1446}

\usepackage{microtype}
\usepackage{enumitem}
\usepackage{amsmath}
\usepackage{framed}
\usepackage{rotating}
\usepackage{pifont}
\usepackage[verbose]{placeins} 
\usepackage{colortbl}
\usepackage{tabu}
\usepackage{arydshln}
\usepackage{subcaption}
\usepackage{adjustbox}
\usepackage{array}
\usepackage{framed}
\usepackage{multirow}
\usepackage{graphicx}
\usepackage{float}
\usepackage{stfloats}
\usepackage[dvipsnames,svgnames]{xcolor}
\usepackage{tcolorbox}
\usepackage{scalerel}
\usepackage{graphicx}
\usepackage{balance}
\usepackage{svg}

\newcommand{\summarytcolorbox}[2]{
\begin{tcolorbox}[% 
  colback=purple!5!white,
  colframe=purple!75!black,
  colbacktitle=purple!60!black, % Darker teal background for the title
  boxrule=0.5pt, % Thickness of the border (can reduce to 0pt if no border needed)
  left=2pt, % Adjust left padding
  right=2pt, % Adjust right padding
  top=2pt, % Adjust top padding
  bottom=2pt, % Adjust bottom padding
  title=\textbf{\textit{#1}}] % Enable breaking across pages
  \noindent #2
\end{tcolorbox}
}

\vgtccategory{Research}

\title{Your Model Is Unfair, Are You Even Aware?\\Inverse Relationship Between Comprehension and Trust in Explainability Visualizations of Biased ML Models}

\author{%
  \authororcid{Zhanna Kaufman}{0000-0002-9135-815X},
  \authororcid{Madeline Endres}{0000-0002-1451-4083},
  \authororcid{Cindy Xiong Bearfield}{0000-0002-1451-4083}, and
  \authororcid{Yuriy Brun}{0000-0003-3027-7986}
}

\authorfooter{
  \item Zhanna Kaufman, Madeline Endres, and Yuriy Brun, University of Massachusetts.
  	E-mail: \{zhannakaufma, mendres, brun\}@cs.umass.edu.
  \item Cindy Xiong Bearfield, Georgia Institute of Technology. E-mail: cxiong@gatech.edu.
}

\abstract{%
Systems relying on ML have become ubiquitous, but so has biased behavior within them. 
Research shows that bias significantly affects stakeholders' trust in systems and how they use them. Further, stakeholders of different backgrounds view and trust the same systems differently. Thus, how ML models' behavior is explained plays a key role in comprehension and trust. %In this paper, 
We %comprehensively 
survey explainability visualizations, creating a taxonomy of design characteristics. We conduct user studies to evaluate five state-of-the-art visualization tools (LIME, SHAP, CP, Anchors, and ELI5) for model explainability, measuring how taxonomy characteristics affect comprehension, bias perception, and trust for non-expert ML users. 
Surprisingly, we find an inverse relationship between comprehension and trust: the better users understand the models, the less they trust them. We investigate the cause and find that this relationship is strongly mediated by bias perception: more comprehensible visualizations increase people's perception of bias, and increased bias perception reduces trust.
We confirm this relationship is causal: Manipulating explainability visualizations to control comprehension, bias perception, and trust, we show that visualization design can significantly ($p<0.001$) increase comprehension, increase perceived bias, and reduce trust. Conversely, reducing perceived model bias, either by improving model fairness or by adjusting visualization design, significantly increases trust even when comprehension remains high.  Our work advances understanding of how comprehension affects trust and systematically investigates visualization's role in facilitating responsible ML applications. 
}


\teaser{
  \centering
  \vstretch{.88}{\includegraphics[trim={0 .21cm 0 .88cm},clip,width=0.90\linewidth]{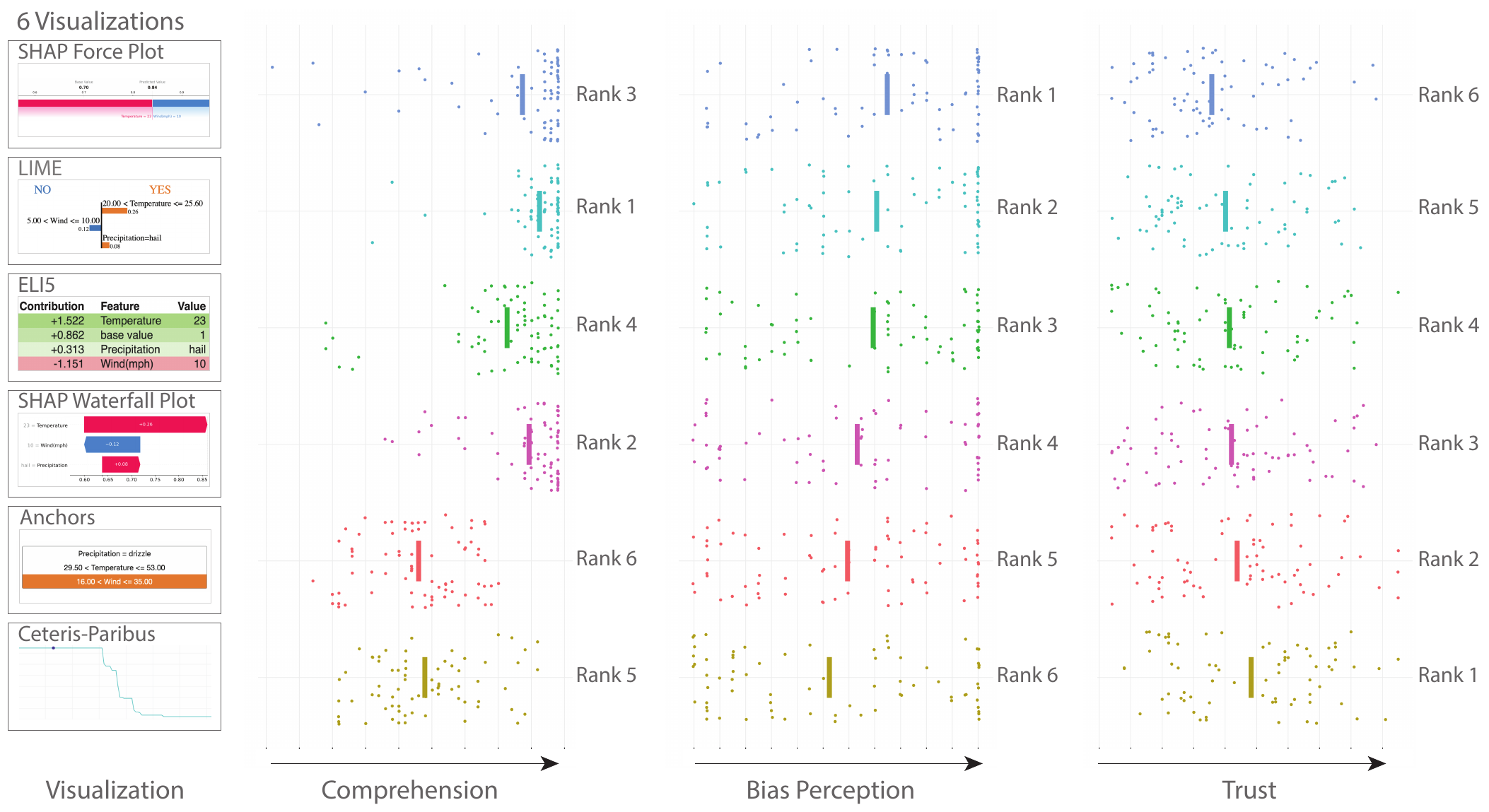}}
  \caption{
  The distributions and means %(with means indicated via vertical bars) 
  of participants' comprehension, bias perception, and trust in ML models across five explainability visualization tools: SHAP, LIME, ELI5, Anchors, and Ceteris-Paribus Profiles. We find a negative correlation between comprehension and trust, strongly mediated by bias perception.  That is, \emph{the better} visualizations explain biased ML models, the less people trust those models. The strong negative correlation between bias perception and trust can be seen by comparing the middle and right graphs.}
  \label{fig:teaser}
}

\renewcommand{\manuscriptnotetxt}{}

\graphicspath{{figs/}{figures/}{pictures/}{images/}{./}} % where to search for the images

\usepackage{tabu}                      % only used for the table example
\usepackage{booktabs}                  % only used for the table example
\usepackage{lipsum}                    % used to generate placeholder text
\usepackage{mwe}                       % used to generate placeholder figures

\usepackage{mathptmx}                  % use matching math font

\begin{document}

%%%%%%%%%%%%%%%%%%%%%%%%%%%%%%%%%%%%%%%%%%%%%%%%%%%%%%%%%%%%%%%%
%%%%%%%%%%%%%%%%%%%%%% START OF THE PAPER %%%%%%%%%%%%%%%%%%%%%%
%%%%%%%%%%%%%%%%%%%%%%%%%%%%%%%%%%%%%%%%%%%%%%%%%%%%%%%%%%%%%%%%


\firstsection{Introduction}
\label{sec:Introduction}

\maketitle

Modern software systems increasingly rely on machine learning (ML), including in high-impact societal domains such as healthcare~\cite{Komorowski18}, hiring~\cite{Roy20}, banking~\cite{qai22}, and criminal justice~\cite{Angwin16}.
Stakeholders ranging from engineers and domain experts to policymakers and end-users routinely make critical decisions about these ML-driven systems~\cite{Hoag23icse-demo, President16, Vanderford22}.
These decisions span from choosing which ML technology to use, to how it is integrated, to which systems consumers trust and buy.  

Unfortunately, %despite ML's increasing ubiquity, 
modern ML systems frequently exhibit sexist, racist, and otherwise biased behavior~\cite{Thomas19science}. 
For example, ML-based systems can have lower cancer detection rates for people of color~\cite{Yaya19}, 
facial recognition systems can discriminate based on sex and race~\cite{Buolamwini18}, and 
software often overestimates recidivism likelihoods for people of color~\cite{Angwin16}. 
Such biases have prompted legal restrictions on certain ML systems~\cite{Singer19, Sheard22}. 
Thus, to make informed decisions, stakeholders need to \emph{understand} model behavior and detect potential \emph{biases}.
It is essential to help non-ML-experts accurately interpret and evaluate ML models.

\emph{Explainability visualizations}, graphical representations clarifying ML model behavior, can help improve understanding~\cite{Cabrera19, Gaba23vis, Johnson23fairkit, Bhatt20, Mertes22, Ozalp23, Heimerl22}. For tabular data classifiers, local explainability visualizations highlight the importance of specific feature values (e.g., how age affects loan rejection). The left column of Figure~\ref{fig:teaser} shows six such visualizations: SHAP force~\cite{Lundberg18} and waterfall plots~\cite{Lundberg17}, LIME bar charts~\cite{Ribeiro16}, ELI5 tables~\cite{Korobov17}, Anchors~\cite{RibeiroTulio18}, and Ceteris-Paribus Profiles~\cite{Biecek18}.

These visualizations facilitate user comprehension~\cite{Ribeiro16, Bansal21} and bias perception~\cite{Berkel21, Ghai22}. However, they are highly heterogeneous in how they indicate feature attribution (e.g., color, positioning, or shape), present alternative input instances, and communicate information (e.g., explicitly or implicitly).
Design choices can profoundly affect reasoning~\cite{Xiong22}, causal conclusions~\cite{Xiong20}, and fairness perceptions~\cite{Berkel21, Gaba23vis, Wang20}. 
Yet, there exists a limited systematic understanding of how visualization design influences comprehension, bias detection, and trust.

To address this need, we systematically explore how visualization design decisions affect stakeholders' ability to comprehend ML model behavior, accurately detect bias, and appropriately allocate trust. 
We focus on local explainability visualizations for tabular classifiers due to their widespread critical use (e.g., in healthcare~\cite{Komorowski18}, hiring~\cite{Roy20}, and finance~\cite{qai22}). 
Our goal is insights that generalize across heterogeneous models, capture perceptions from non-ML-expert users, and capture causality. 
These insights will enable the design of visualizations that align perception of bias with actual model behavior. 

Our approach leverages two key insights. First, systematic study of state-of-the-art tools allows us to build a taxonomy capturing abstract design variability  across visualizations, highlighting characteristics critical for comprehension and perception. Second, abstracted taxonomy features enable controlled evaluation to causally examine how visualization characteristics impact comprehension, bias detection, and trust, generating insights likely to generalize beyond individual tools.

\looseness-1
We find that visualization design significantly influences comprehension, trust, and bias perception. 
Providing explicit feature attribution values significantly increases comprehension and bias detection.
Simplifying visualizations (reducing the number of visual characteristics) also increases bias perception, even when comprehension remains constant. 

Surprisingly, visualizations that increase comprehension, reduce trust. 
We investigate the cause of this relationship and find that it is strongly mediated by bias perception($p < 0.001$ for all): more comprehensible visualizations may increase bias perception, and a higher perception of bias reduces trust. 

We also demonstrate that specific design choices (e.g., implicit vs.\ explicit feature values) can enable more accurate detection of bias ($d=0.22$, $p<0.001$). Reducing actual model bias increases trust ($d=0.44$, $p<0.001$), but decreasing bias perception without altering actual bias and minimally impacting comprehension %($d=-0.065$, $p=0.01$) 
can still increase trust ($d=-0.19$, $p<0.001$).

This paper makes the following contributions:

\begin{itemize}[itemsep=0pt, topsep=0pt, leftmargin=10pt]

\item A systematic review of 26 visualizations, producing a \emph{comprehensive taxonomy of visualization design characteristics}, abstracted from state-of-the-art local explainability visualization tools.

\item A series of {\emph{large-scale user studies} with 818 non-expert participants, empirically assessing how design characteristics correlate with and causally influence comprehension, bias perception, and trust.}

\end{itemize}

\noindent Our stimuli and data are available in our supplementary materials~\cite{supp}.

\section{Background: ML Explanation Visualizations}
\label{sec:Related Work}

As ML use has become more common, the importance of non-experts understanding ML model behavior %when making decisions about both system design and use 
has increased.  
Non-experts often need to understand and use ML models~\cite{bagrow20, Salehin24, Yang18} and non-experts' input can improve ML automation~\cite{Gennatas20, Nakao20, Crandall18}, creating a need for effective explainability tools~\cite{Bhatt20} that target non-experts~\cite{Bhatt20, Mertes22, Ozalp23, Heimerl22}. 

The needs for ML explainability are significant.  
End-users often have difficulty understanding ML decisions~\cite{Hartwig22}, which can deteriorate their trust and use of ML tools~\cite{Diprose20}.
The European Union's General Data Protection Regulation even requires subjects of decisions made by ML systems to have a right to an explanation~\cite{GDPR2016a}.
Additionally, system designers need to understand ML models when incorporating them into their systems.  
In practice, however, many ML models, including deep learning, are so complex that even ML experts struggle to understand their functionality~\cite{Samek17}. 
Debugging unexpected behavior, particularly when that behavior is driven by complex, opaque ML models, can be labor-intensive %, meticulous,
and slow~\cite{Liu24}. 
Biased model behavior, which is unfortunately common in ML~\cite{Galhotra17fse, Thomas19science, Veale17} adds an extra layer of complexity; bias can be particularly difficult to formalize and control during training because of a possible trade-off between fairness and accuracy~\cite{Liang23}, and different notions of bias can be incompatible~\cite{Friedler16}.

Visualization offers hope.
Visualizing model training %processes 
can support debugging~\cite{Liu24}, and improve explainability for hard-to-formalize properties, such as safety~\cite{Burkart21}.
It can also clarify behavior when training and deployment environments differ~\cite{Lipton17}, and support the detection and debugging of bias~\cite{Ahn19, Johnson23fairkit, Johnson20icse}. 
However, few insights exist into how visualization design choices influence comprehension, bias perception, and trust. 
Providing these insights is the goal of this paper.

\subsection{ML Explainability}
\label{sec:ML Explainability}

This paper focuses on the most common type of ML models: classification models for tabular data.
We use the term \emph{model} to refer to classification models %, or classifiers,   
whose goal is classifying inputs into categories, e.g., images of medical tissue scans into cancerous or non-cancerous. 
More formally, an ML model is a mathematical function mapping feature sets to categorical labels.  
ML explainability methods generally fall into three categories: inherently interpretable, global, and local.

\emph{Inherently interpretable} (white-box) models are self-explanatory by design~\cite{Baniecki21},
but often fail to capture intricate feature interactions. 
By contrast, many high-performing models are black-box, requiring model-agnostic approaches to 
explain behavior~\cite{Ribeiro16, Lundberg17,Bastiani19, Hailesilassie16, Blanco19}. 
\emph{Global explanations} use approximations to summarize model behavior across all predictions~\cite{Baniecki21, Hailesilassie16, Yao19, Wang221, Islam21, Ribeiro16, Lundberg17, Staniak18}, offering useful high-level insights. 
However, they often lack the granularity needed to %understand specific outputs, especially 
assess fairness or trust in individual predictions. 

\emph{Local explanations} describe model behavior for individual input instances~\cite{Ribeiro16, Lundberg17, Staniak18}. 
There are many local explanation types, which vary substantially in visual design and emphasis. 
Feature importance scores quantify how each feature impacts the model's predicted probabilities~\cite{Wang221}.
Ceteris-Paribus~(CP) profiles, for each input, graph changes in predicted model output (y axis) as a result of changing a single feature in isolation (x axis)~\cite{Biecek21}. 
Individual conditional expectation~(ICE) plots combine multiple CP profiles for the same feature, each representing a different input~\cite{Goldstein15}. 
Contrastive explanations compare model outputs when feature values change~\cite{Stepin24}. 
Counterfactual explanations 
identify feature-value modifications that alter predictions~\cite{Wexler19, Guidotti22}.
Finally, pertinent negatives find the minimal feature changes needed to flip model predictions, while pertinent positives show the smallest feature subset that must remain fixed to maintain the same prediction~\cite{Klaise21}. 

This paper investigates how local explanation visualizations facilitate comprehension and trust among non-expert ML users, particularly when models are unfair. 
We focus on local explanations because they provide  granular logic for model behavior: a global explanation may highlight important features, but local explanations reveal how individual feature values shift model predictions. 
Local explanations require the user to infer broader patterns from individual examples~\cite{Wielopolski2024}, making them especially relevant for understanding bias perception and trust. 
We focus on tabular data classifiers because of their significant design heterogeneity. This provides a rich opportunity to examine how design influences comprehension and trust, particularly for unfair models. Specialized explanations for image and text classifiers are often visually similar across tools~\cite{Wang221} (e.g., highlighting salient words or pixels). 

While several explainability tools focus specifically on revealing bias during or after model development~\cite{Johnson23fairkit, Johnson22, Ahn19, Cabrera19, Xie22, Ghai22, Wang21}, they target programmers and data scientists. 
They require ML and bias expertise, include complex interfaces, and require prior understanding of model functionality. By contrast, we target a broader range of users, including non-experts. 
We thus focus on simpler tools that explain model behavior without requiring specialized background knowledge. 

\subsection{Explainability, Comprehension, and Trust}
\label{sec:trustbackground}

We now overview current understanding of user perceptions of visualization explanations, focusing on model comprehension and trust.

\textbf{Explainability and Comprehension.} Assessing comprehension is essential for understanding explainability visualization effectiveness because comprehension is a measure of model interpretability. However, comprehension remains surprisingly understudied;  only 22\% of  AI explainability tools include user studies measuring model comprehension~\cite{Nauta23}. Common methods include (1)~asking participants to compare features' predictive power~\cite{Collaris21}, predict model outputs~\cite{Yao19, Colin23}, or match a prediction to an explanation~\cite{Kim22}, (2)~measuring how often users made correct decisions when advised by a model~\cite{Hoque22, Shen20, Alufaisan20} or detecting incorrect predictions~\cite{Kim22}, and (3)~using open-ended questions about model decisions~\cite{Ribeiro16}. 
Generally, these studies suggest that explanations improve comprehension. 
However, due to the wide variety of tool designs and evaluation methods, comparing comprehension across different explanation visualizations remains challenging.

In this work, we integrate these approaches into a metric that 
allows comprehension comparisons across  explainability visualizations. We define comprehension as an understanding of the model's output for a given input, along with the magnitude and direction of each input feature towards a specific classification outcome.

\textbf{Explainability and Trust.}  The most common measure used for ML model \emph{trust}---willingness to accept a model's outputs---often does not reflect people's actual trust levels~\cite{Ahn24}, particularly when unfairness is involved. 
People may accept a model's recommendation, but distrust it due to awareness of broader socio-organizational contexts%for that model's decisions
~\cite{Ehsan21}. 
Furthermore, users may prefer AI recommendations over human ones, even when perceiving the AI less morally trustworthy~\cite{Tolmeijer22}. 
By contrast, we define trust as a combination of perceived accuracy and willingness to rely on a model for decisions affecting users and others.

Prior studies report mixed findings on interpretability's effect on trust. 
Decision explanations can increase trust in model accuracy~\cite{Ribeiro16,Yang20}, and  bias-focused explainability tools may foster trust~\cite{Ghai22}. 
In some domains, the mere presence of explanations can make people more likely to use an ML model~\cite{Bansal21}, while in others, experts use their prior knowledge to calibrate their trust in a model based on its explanation~\cite{Wang22}. 
Conversely, observed accuracy can have a more significant effect than interpretability~\cite{Ahn24} on both user trust and willingness to use decision-making assistants~\cite{Rechkemmer22, Yin19, Rechkemmer22, Zhang20}. 
Increased comprehension does not necessarily promote  model use~\cite{Sangdeh21}; 
practitioners 
can misunderstand and over-trust visualizations, mistakenly dismissing suspicious results~\cite{Kaur20},  leading to misguided decisions~\cite{Ma23, He23}.
Finally, perceptions of explainability methods can vary by audience demographics, affecting trust and bias perception~\cite{Arya19, Gaba23vis, Lee21}.

Visualization design shapes both people's perception of data and their decision-making strategies~\cite{Xiong22, Xiong20, Wang20}. 
Different visualization types affect trust, with familiar, simpler, or aesthetically pleasing  visualizations increasing trust the most~\cite{Crouser24, Elhamdadi24}.  Visualization design can impact emotional reactions, driving important decisions~\cite{Yang24}. 
It can also reveal model biases~\cite{Berkel21} and affect reactions to biased models~\cite{Gaba23vis}. 
This paper fills the need for further research into how specific design characteristics affect the relationship between comprehension, bias perception, and trust. %interpretability (as measured by comprehension, explainability visualizations, and trust). 
We perform this investigation using the six state-of-the-art explainability visualizations seen in Figure~\ref{fig:vis-combined}.

\section{Research Questions}
\label{sec:RQs}

We organize our investigation of how explainability visualizations can impact comprehension and trust around three research questions:

\begin{itemize}[leftmargin=26pt,itemsep=0pt, topsep=0pt]

  \item[\textbf{RQ1:}] What design characteristics of ML explainability visualizations could impact comprehension, trust, and bias perception?
    
  \item[\textbf{RQ2:}] Do visualization design characteristic variations 
  %across state-of-the-art tools 
  correlate with differences in model comprehension, bias perception, and trust? 

  \item[\textbf{RQ3:}] Are the observed relationships between comprehension, bias perception, and trust causal? Do the relationships generalize beyond existing tools?
  
\end{itemize}

We address RQ1 via the construction of a visualization characteristic taxonomy (Section~\ref{sec:Taxonomy}). We then design and conduct user studies (Section~\ref{sec:Methodology}) to address RQ2 (Section~\ref{sec:RQ2}) and RQ3 (Section~\ref{sec:RQ3}).

\section{RQ1: What Visualization Characteristics Exist?}
\label{sec:Taxonomy}

We want to identify visualization design characteristics that could impact model comprehension and trust. We desire an understanding that (1)~covers state-of-the-art tools and (2)~abstracts key characteristics that can be varied experimentally. 
We systematically analyze existing explanation visualization tools to construct a comprehensive taxonomy.
 
\subsection{Taxonomy Development Methodology}
\label{sec:Taxonomy Development Methodology}

To build our taxonomy, we first compile a representative collection of state-of-the-art local explanation visualization tools. 
We start with four recent surveys of visualization tools~\cite{Maksymiuk21, Burkart21, Nauta23, Rong24} (including one from 2024), supplemented by an updated list of explainability resources~\cite{Hall24}. 
From these sources, we select all tools with publicly available implementations (enabling practical evaluation), excluding tools that provide purely textual explanations, non-local explanations, and explanations for non-tabular data. Through this method, we collect a set of 26 visualizations from 20 visualization-based ML explainability tools, which we empirically analyze using the taxonomy development methodology described by Nickerson et al.~\cite{Nickerson2013AMF}.

Nickerson et al.~\cite{Nickerson2013AMF} define a taxonomy as a set of $n$ dimensions, each consisting of $k$ mutually exclusive characteristics, such that every classified member has exactly one characteristic per dimension. Following their empirical approach, we begin by defining our \emph{meta-characteristic}, a central purpose from which all other characteristics derive. Using our comprehension definition from Section~\ref{sec:trustbackground}, we derive our meta-characteristic:
\emph{How does a visualization convey information about the impact of each individual feature on a single model prediction?}

We follow Nickerson et al.'s suggested iterative methodology. The first author analyzes progressively larger subsets of visualizations to identify characteristics related to our meta-characteristic (e.g., color, position, graph, etc.), consulting with co-authors at intermediate steps to establish consensus. These characteristics are then grouped into dimensions, such that each  visualization has exactly one characteristic per dimension.
This iterative process continues until the dimensions comprehensively reflect our meta-characteristic.  

\subsection{A Taxonomy of Visualization Design Characteristics}

\newcolumntype{R}[2]{%
    % >{\adjustbox{angle=#1,lap=\width-(#2)}\bgroup}%
    >{\adjustbox{angle=#1,lap=(#2)-\width}\bgroup}%
    p{1em}%
    <{\egroup}%
}
\newcommand*\rot{\multicolumn{1}{R{25}{1em}}}% no optional argument here, please!

\begin{figure}[t!!!]
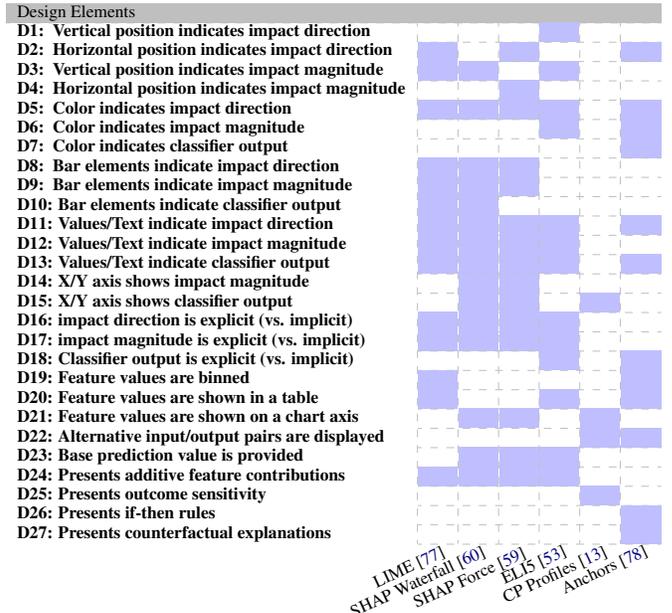

\centerline{%
\resizebox{\columnwidth}{!}{
% \small
% \centering
\taburulecolor{lightgray}
\begin{tabular}{l:c:c:c:c:c:c} 
\rowcolor{gray!50}\multicolumn{7}{l}{Design Elements}\\ \cdashline{1-7}
{\textbf{D1:~~Vertical position indicates impact direction}} & & & & \cellcolor{blue!25} & & \\ \cdashline{2-7}
{\textbf{D2:~~Horizontal position indicates impact direction}} & \cellcolor{blue!25} & & \cellcolor{blue!25} & & & \cellcolor{blue!25} \\ \cdashline{2-7}
{\textbf{D3:~~Vertical position indicates impact magnitude}} & 
\cellcolor{blue!25} & \cellcolor{blue!25} & & \cellcolor{blue!25} & & \\ \cdashline{2-7}
{\textbf{D4:~~Horizontal position indicates impact magnitude}} & 
& & \cellcolor{blue!25} & & & \\ \cdashline{2-7}
{\textbf{D5:~~Color indicates impact direction}} & 
\cellcolor{blue!25} & \cellcolor{blue!25} & \cellcolor{blue!25} & \cellcolor{blue!25} & & \cellcolor{blue!25} \\ \cdashline{2-7}
{\textbf{D6:~~Color indicates impact magnitude}} & & & & \cellcolor{blue!25} & & \cellcolor{blue!25} \\ \cdashline{2-7}
{\textbf{D7:~~Color indicates classifier output}} &
& & & & & \cellcolor{blue!25} \\ \cdashline{2-7}
{\textbf{D8:~~Bar elements indicate impact direction}} & \cellcolor{blue!25} & \cellcolor{blue!25} & \cellcolor{blue!25} & & & \\ \cdashline{2-7}
{\textbf{D9:~~Bar elements indicate impact magnitude}} & \cellcolor{blue!25} & \cellcolor{blue!25} & \cellcolor{blue!25} & & & \\ \cdashline{2-7}
{\textbf{D10: Bar elements indicate classifier output}} & 
\cellcolor{blue!25} &
\cellcolor{blue!25} & & & & \\ \cdashline{2-7}
{\textbf{D11: Values/Text indicate impact direction}} &
\cellcolor{blue!25} & \cellcolor{blue!25} & \cellcolor{blue!25} & \cellcolor{blue!25} & & \cellcolor{blue!25} \\ \cdashline{2-7}
{\textbf{D12: Values/Text indicate impact magnitude}} & \cellcolor{blue!25} & \cellcolor{blue!25} & \cellcolor{blue!25} & \cellcolor{blue!25} & & \\ \cdashline{2-7}
{\textbf{D13: Values/Text indicate classifier output}} &
\cellcolor{blue!25} & \cellcolor{blue!25} & \cellcolor{blue!25} & \cellcolor{blue!25} & & \cellcolor{blue!25} \\ \cdashline{2-7}
{\textbf{D14: X/Y axis shows impact magnitude}} 
& & \cellcolor{blue!25} & \cellcolor{blue!25} & & & \\ \cdashline{2-7}
{\textbf{D15: X/Y axis shows classifier output}}
& & \cellcolor{blue!25} & \cellcolor{blue!25} & & \cellcolor{blue!25} & \\ \cdashline{2-7}
{\textbf{D16: impact direction is explicit (vs. implicit)}} &
\cellcolor{blue!25} & \cellcolor{blue!25} & \cellcolor{blue!25} & \cellcolor{blue!25} & & \\ \cdashline{2-7}
{\textbf{D17: impact magnitude is explicit (vs. implicit)}} &
\cellcolor{blue!25} & \cellcolor{blue!25} & \cellcolor{blue!25} & \cellcolor{blue!25} & & \\ \cdashline{2-7}
{\textbf{D18: Classifier output is explicit (vs. implicit)}} &
& & & \cellcolor{blue!25} & & \cellcolor{blue!25} \\ \cdashline{2-7}
{\textbf{D19: Feature values are binned}} & 
\cellcolor{blue!25} & & & & & \cellcolor{blue!25} \\ \cdashline{2-7}
{\textbf{D20: Feature values are shown in a table}} & 
\cellcolor{blue!25} & & & \cellcolor{blue!25} & & \cellcolor{blue!25} \\ \cdashline{2-7}
{\textbf{D21: Feature values are shown on a chart axis}} & 
& \cellcolor{blue!25} & \cellcolor{blue!25} & & \cellcolor{blue!25} & \\ \cdashline{2-7}
{\textbf{D22: Alternative input/output pairs are displayed}} & & & & & \cellcolor{blue!25} & \cellcolor{blue!25} \\ \cdashline{2-7}
{\textbf{D23: Base prediction value is provided}} & 
& \cellcolor{blue!25} & \cellcolor{blue!25} & \cellcolor{blue!25} & & \\ \cdashline{2-7}
{\textbf{D24: Presents additive feature contributions}} & \cellcolor{blue!25}
& \cellcolor{blue!25} & \cellcolor{blue!25} & \cellcolor{blue!25} & & \\ \cdashline{2-7}
{\textbf{D25: Presents outcome sensitivity}} & 
& & & & \cellcolor{blue!25} & \\ \cdashline{2-7}
{\textbf{D26: Presents if-then rules}} & 
& & & & & \cellcolor{blue!25} \\ \cdashline{2-7}
{\textbf{D27: Presents counterfactual explanations}} & & & & & & \cellcolor{blue!25} \\ \cdashline{2-7}
% \multicolumn{1}{c}{} & \rot{LIME~\cite{Ribeiro16}} & \rot{SHAP Waterfall~\cite{Lundberg17}} & \multicolumn{1}{c}{\rotatebox{70}{SHAP Force~\cite{Lundberg18}}} & \multicolumn{1}{c}{\rotatebox{70}{ELI5~\cite{eli5}}} & \multicolumn{1}{c}{\rotatebox{70}{Ceteris-paribus (CP) Profiles~\cite{Biecek21}}} & \multicolumn{1}{c}{\rotatebox{70}{Breakdown (Dalex)~\cite{Staniak18}}} & \multicolumn{1}{c}{\rotatebox{70}{iBreakdown~\cite{Gosiewska20}}} & \multicolumn{1}{c}{\rotatebox{70}{Breakdown Violin Plot (Dalex)~\cite{Biecek18}}} & \multicolumn{1}{c}{\rotatebox{70}{Shapley(ModelStudio)~\cite{Biecek18}}} & \multicolumn{1}{c}{\rotatebox{70}{LIME (Dalex)~\cite{Biecek21}}} & \multicolumn{1}{c}{\rotatebox{70}{iml LIME (Dalex)~\cite{Biecek18}}} & \multicolumn{1}{c}{\rotatebox{70}{InterpretML~\cite{Nori19}}} & \multicolumn{1}{c}{\rotatebox{70}{ExplainX~\cite{explainx}}} & \multicolumn{1}{c}{\rotatebox{70}{shapper (Dalex)~\cite{Biecek21}}} & \multicolumn{1}{c}{\rotatebox{70}{SHAP (ModelOriented) ~\cite{Biecek18}}} & \multicolumn{1}{c}{\rotatebox{70}{Perturbation feature importance (ModelDown)~\cite{Biecek18} }} & \multicolumn{1}{c}{\rotatebox{70}{Integrated Gradients(Alibi)~\cite{Klaise21}}} & \multicolumn{1}{c}{\rotatebox{70}{What-if Tool~\cite{Wexler19}}} & \multicolumn{1}{c}{\rotatebox{70}{Counterfactual Instances (Alibi)~\cite{Klaise21}}} & \multicolumn{1}{c}{\rotatebox{70}{Pertinent Negatives/Positives (Alibi)~\cite{Klaise21}}} & \multicolumn{1}{c}{\rotatebox{70}{Vertex AI (Google)~\cite{vertex}}} & \\
\multicolumn{1}{c}{} & 
\rot{~~~~~~~~~~~~~~~~~~~~~~~~~~~~~~~~~~~~~~~~~~~~~~~~~~~~~~~~~~~~~~~~~~~~LIME~\cite{Ribeiro16}} & 
\rot{~~~~~~~~~~~~~~~~~~~~~~~~~~~~~~~~~~~~~~~~~~~~~~~~~~~~SHAP Waterfall~\cite{Lundberg17}} & 
\rot{~~~~~~~~~~~~~~~~~~~~~~~~~~~~~~~~~~~~~~~~~~~~~~~~~~~~~~~~~SHAP Force~\cite{Lundberg18}} & 
\rot{~~~~~~~~~~~~~~~~~~~~~~~~~~~~~~~~~~~~~~~~~~~~~~~~~~~~~~~~~~~~~~~~~~~~~~ELI5~\cite{Korobov17}} & 
\rot{~~~~~~~~~~~~~~~~~~~~~~~~~~~~~~~~~~~~~~~~~~~~~~~~~~~~~~~~~~~~CP Profiles~\cite{Biecek21}} & 
\rot{~~~~~~~~~~~~~~~~~~~~~~~~~~~~~~~~~~~~~~~~~~~~~~~~~~~~~~~~~~~~~~~~~Anchors~\cite{RibeiroTulio18}}\\     
\end{tabular}
} % resizebox
} % centerline
\vspace{-12ex}
\caption{Our visualization design characteristic taxonomy, with 54 characteristics across 27 dimensions (D1--D27), on the six visualizations used in our user study (Section~\ref{sec:visualization choices}). Shaded cells indicate that a specific dimension was characterized as true, while unshaded cells indicate it was characterized as false. 
Our supplementary materials~\cite{supp} include the complete taxonomy applied to all 26 analyzed visualizations.}
\label{fig:elements in tools}
\end{figure}

Figure~\ref{fig:elements in tools} shows the 54 characteristics we found, grouped into 27 dimensions, D1--D27, with 2 characteristics per dimension (true/false). This figure also shows our encoding for the explainability visualizations in Figure~\ref{fig:vis-combined}. The dimensions are summarized here:

\begin{itemize}[labelwidth=0.7em, labelsep=0.6em, topsep=0ex, itemsep=0ex,
  parsep=0ex, leftmargin=1.5em]

\item Visualization element \emph{colors} convey the direction~\textbf{(D5)} or magnitude~\textbf{(D6)} of feature impact, or the value of the prediction~\textbf{(D7)}.

  \item \emph{Bar elements} in the visualization convey the direction~\textbf{(D8)} or magnitude~\textbf{(D9)} of feature impact, or the value of the prediction~\textbf{(D10)}.

  \item \emph{Printed numerical values} show direction~\textbf{(D11)} or magnitude~\textbf{(D12)} of feature impact, or prediction values~\textbf{(D13)}.

  \item Magnitude of feature impact~\textbf{(D14)} or the value of the prediction~\textbf{(D15)}, are on a numerical \emph{x or y axis}. 

  \item Feature impact direction~\textbf{(D16)} and magnitude~\textbf{(D17)}, or prediction values~\textbf{(D18)} are conveyed \emph{explicitly} or \emph{implicitly}. 

  \item Feature impacts are specific to individual values or generalized within value \emph{bins}~\textbf{(D19)}.

  \item Individual input \emph{feature values are located} in a separate table, or indicated on the axis of a graph element~\textbf{(D20--D21)}. 

  \item Visualization may provide \emph{alternative example feature inputs} and resulting predictions~\textbf{(D22)}.

  %\item Visualization may provide a \emph{bias/base value} influencing predictions~\textbf{(D22)}. 
  \item Visualization may provide a \emph{bias/base value} for predictions~\textbf{(D23)}.

\item Visualizations present different explanation types, including \emph{Additive feature contributions, outcome sensitivity to input changes, counterfactual explanations}, and \emph{if-then rules}~\textbf{(D24--D27)}. 
  %that is acted on by the input features to produce the final prediction output. 

\end{itemize}

Our taxonomy of visualization design characteristics enables both evaluating existing explainability visualizations and designing new methods of conveying this information. For example in RQ3, we show how characteristics such as explicit indications of impact magnitude and classifier output (D17 and D18) can impact user comprehension and bias perception of the underlying model.
 
\summarytcolorbox{RQ1 Summary: Explainability Visualization Characteristics}{Our taxonomy  analysis identified 44 characteristics of visualization design characteristics
across 26 ML explainability visualizations from 20 visualization tools intended to promote comprehension of tabular data classifiers (Figure~\ref{fig:elements in tools}). These characteristics allow us to systematically assess the impact of design decisions on user comprehension and perception of explainability visualizations.
} 

\section{RQ2 and RQ3 User Study Design}
\label{sec:Methodology}

\begin{figure}[t]
% \small

% \begin{center}
\resizebox{\columnwidth}{!}{
\begin{tabular}{@{}llllcr@{}}
\toprule
Metric & Type & No. & Text \\ 
\midrule
\multirow{3}{*}{\shortstack[l]{Compre- \\ hension}} & \multirow{3}{*}{\shortstack[l]{Multiple \\ Choice}} & C1 & Will this model approve the loan for this person? \\ 
\cmidrule{3-6}
 & & C2 & What feature has the most predictive power for this decision? \\ 
\cmidrule{3-6}
 & & C3 & Which factor(s) are pushing the model toward predicting `NO'/`YES' \\ 
\midrule
\multirow{3}{*}{\shortstack[l]{Perceived \\ Compre- \\ hension}} & \multirow{3}{*}{\shortstack[l]{Likert}} & PC1 & How well did you understand the way this model makes decisions? \\ 
\cmidrule{3-6}
 & & PC2 & How easy was it for you to understand the model output? \\ 
\cmidrule{3-6}
 & & PC3 & How likely would you use this visualization to explain models to other people? \\ 
\midrule
\multirow{4}{*}{Trust} & \multirow{4}{*}{Likert} & & On a scale of 1 to 6, how much do you trust the model to approve or deny a loan \ldots \\ 
\cmidrule{3-6}
 & & T1 & \shortstack[l]{\ldots for you?} \\ 
\cmidrule{3-6}
 & & T2 & \shortstack[l]{\ldots for other people in general?} \\ 
\cmidrule{3-6}
 & & T3 & I trust the data this model was trained on. \\ 
\cmidrule{3-6}
 & & T4 & This model is accurate. \\ 
\cmidrule{3-6}
 & & T5 & Computer models can be trusted to make human decisions. \\ 
\midrule
\multirow{4}{*}{\shortstack[l]{Bias \\ Perception}} & \multirow{4}{*}{Yes/No} & B1 & Do you think this model includes potentially discriminating factors? \\ 
\cmidrule{3-6}
 & & B2 & This model uses all of the features that it should use when making this decision. \\ 
\cmidrule{3-6}
 & & B3 & This model does not use any unnecessary features when making this decision. \\ 
\cmidrule{3-6}
 & & B4 & This model is fair. \\ 
\midrule
\multirow{5}{*}{\shortstack[l]{Behavioral \\ Alignment}} & \multirow{5}{*}{Yes/No} & & This model would probably give me a loan \ldots \\ 
\cmidrule{3-6}
 & & A1 & \ldots because I am similar to the person described in this question. \\ 
\cmidrule{3-6}
 & & A2 & \ldots because I am different from the person described in this question. \\ 
\cmidrule{3-6}
 & & A3 & \ldots because of previous decisions it has made. \\ 
\cmidrule{3-6}
 & & A4 & This model would probably not give me a loan, and this would be the correct decision.$\!\!\!\!$ \\ 
\bottomrule
\end{tabular}
}
\caption{Overview of survey questions for our three metrics---Comprehension, Trust, Bias Perception---along with questions assessing if participants perceive that model behavior aligns with their expectations. }
%\caption{\todo{This figure needs a lot of work.} 
%\todo{This caption is not good.  It should read: ``The four metrics --- Comprehension, Trust, Bias Perception, and Qualitative --- each consists of multiple questions.}
%\todo{why are some questions numbered Q and others S? Why is this not explained?}
%A table describing questions for each metric. Here you can see the metric, the question type, the question number, the question text, the number of times the question is encountered, and the maximum score value a participant can receive for each question. 
%\todo{where is * Seen described in the text? or the caption?} \todo{change ``Cap'' to ``max'' What is cap? The max value doesn't really make sense to include here.  Why are we telling them this number?  Who needs to know this?  Do we ever report values for these?  Maybe they should be \% instead if we report them?  Rethink * Seen label.  Rethink this table.}}
\label{fig:measures}
% \end{center}
\end{figure}

We next investigate how visualization design elements influence model comprehension, bias perception, and trust. To do so we conducted a series of online Qualtrics~\cite{Qualtrics13} surveys with crowd-sourced participants from Prolific.com~\cite{Palan18}. All our studies were ethics-board approved.

Each survey follows the same format; participants are shown a series of %model inputs, predictions, and 
explainability visualizations for a classifier that recommends whether to give a loan applicant a loan based on various demographic features (e.g., age, education, sex, etc.). For each loan applicant, participants respond to multiple-choice questions designed to assess model comprehension, bias perception, and trust. The surveys differ in either the explanation visualization used or the underlying model and its fairness. 

In total, we conducted eleven surveys: six for RQ2 and five for RQ3. For RQ2, since we are interested in understanding if visualization design characteristics across state-of-the art tools correlate with model comprehension, bias perception, and trust, we present participants with explainability visualizations generated by one of six state-of-the-art tools, carefully chosen due to their high coverage of taxonomic characteristics (see Section~\ref{sec:visualization choices}). 

For RQ3, as we wish to understand if the observed relationships are causal and if the impact of design characteristics on comprehension, bias perception, and trust generalize beyond existing tools, we use five survey variations with altered explainability visualizations to conduct \textit{three crowd-sourced follow-up experiments designed for controlled} analysis of causality (see Section~\ref{sec:RQ3}): 
\begin{itemize}[labelwidth=0.7em, labelsep=0.6em, topsep=1pt, itemsep=2pt,
  parsep=0ex, leftmargin=1.5em]

\item \emph{Experiment 1---Explicitness:} Increase the explicitness of an implicit visualization to modulate comprehension and confirm a causal effect on bias perception. 

\item \emph{Experiment 2---Fairness:}  Increase fairness of the underlying model to modulate bias perception and confirm a causal effect on trust. 

\item \emph{Experiment 3---Bias Perception:} Manipulate design characteristics to alter bias perception while keeping comprehension high to confirm bias perception as the mediating factor.

\end{itemize}

Section~\ref{sec:Measures} details our metrics for measuring comprehension, bias perception, and trust;  Section~\ref{sec:ML model and its instances} describes our explainability visualization stimuli; and Section~\ref{sec:Participants} summarizes our participants.

\subsection{Experimental Measurements}
\label{sec:Measures}

Using a series of multiple-choice questions after each explanation visualization scenario, we measured three primary aspects: comprehension of the underlying ML model, perceived bias in the model, and subsequent trust of that model. We also assessed behavioral alignment to investigate if people's perception of model behavior correlated with their trust. 
We developed our metrics by reviewing measurement methodologies in existing literature.
In addition, we ran two pilot studies, 
each with over 200 participants, to assess our preliminary questions and finalize our metrics. We now discuss each metric in detail.

\emph{Measuring Comprehension:}

We use the definition of comprehension described in Section~\ref{sec:trustbackground}, operationalized via questions C1--C3 in Figure~\ref{fig:measures}. We define an aggregate comprehension score as the sum of correct responses to these questions across all prediction instances. We further measure \textit{perceived comprehension} using three Likert-style questions (PC1--PC3) which assess if people's perception of explainability visualization effectiveness matches their observed comprehension level.

\emph{Measuring Trust:}
We define trust in Section~\ref{sec:trustbackground} as a person's perception of the accuracy of an underlying model and their willingness to rely on the model for decisions that affect themselves and others. We operationalize this definition via the questions T1--T5 in Figure~\ref{fig:measures}, which participants answer for every visualization instance.
We define the aggregate trust score as the sum across all 7 prediction instances.

\emph{Measuring Bias Perception:}
Bias perception measurement tends to be either qualitative or self-reported (e.g., asking participants to identify systemic unfairness~\cite{Nakao22}, rate their general perception of fairness~\cite{Nakao22}, or determine if certain subgroups are treated unfairly~\cite{Collaris21, Wang21}). 
We operationalize perceived bias as a positive response to B1 in Figure~\ref{fig:measures}, combined with the sum of disagreement with statements B2--B4.

\emph{Behavioral Alignment:} People can trust a model more when its decisions benefit them~\cite{Gaba23vis}. 
We summarize this potential behavioral alignment with the Yes/No questions A1--A3 in Figure~\ref{fig:measures}. We also add A4, allowing participants to indicate that the model's decisions would not be beneficial to them, but they still approve of model behavior. 

\emph{Qualitative Analysis:}
We asked participants two free response questions regarding which visualization characteristics they found most and least useful when answering questions about model behavior and model fairness. This allows us additional insights into why certain design characteristics may be associated with higher comprehension or trust. 

\emph{Statistical Methods}
We conduct our analysis in RStudio~\cite{RStudio}. To assess relationships between aggregate scores across visualizations in RQ2 we use linear regression. We use Analysis of Variance (Anova) to determine whether our independent variables are significant predictors of our outcomes, and estimated marginal means (emmeans) to compare average scores for our measures across visualizations. To compare measures across pairs of surveys in RQ3, because our distributions are not normal, we use Wilcoxon Rank Sum Tests. We consider the results significant if $p<0.05$. For effect size, we use Cohen's $d$. Our full analysis can be found in our supplementary materials~\cite{supp}. 

\subsection{Explainability Visualization Stimuli}
\label{sec:ML model and its instances}

Each survey presented participants with 7 scenarios featuring a loan-recommending model. Each scenario consisted of an input (a person wanting a loan), the model's recommendation (approve or deny the loan), and an explanation visualization. 
The model was a LightGBM classifier~\cite{Microsoft24} trained on the Census Income dataset (14~demographic features and income for 48,842 people)~\cite{Becker96}. 
LightGBM classifiers are black-box models that use a gradient boosting decision tree algorithm~\cite{Ke17}.
To reduce visualization complexity, we used a subset of 5 features: age, education level, occupation, hours worked per week, and sex\footnote{We use the term sex and the categories male and female, as in the dataset.}). The model predicts income given these features.
If actual income exceeds the prediction, the model recommends granting the loan.
Because this dataset gives loans to 31\% of men but only 11\% of women, training on it without fairness constraints results in a model that is more likely to recommend a loan to men than to women.

We therefore focused on sex as the model's discriminatory factor, and included 3 visualizations for males and 3 for females, as well as a juxtaposition visualization for a male and a female with equivalent values for all features except sex. 
We chose the applicants (all actual data points from the Census Income dataset) to include a range of education, occupations, and hours worked per week. 
For the two juxtaposed applicants, we filtered the dataset for instances identical in every feature except sex, but for which the model produced different recommendations (one applicant received a loan and the other did not). 

While each survey contained the same seven scenarios, the visualization differed. For RQ2, we consider six state-of-the art explanation visualizations with high coverage of our taxonomy, For RQ3, we consider five additional visualization variations, centered around three experiments testing for causality (see Section~\ref{sec:Methodology}). We detail the RQ2 visualizations in Section \ref{sec:visualization choices}) and the RQ3  variations in Section~\ref{sec:RQ3Methods}.

\subsubsection{State-of-the-Art Explainability Visualizations (for RQ2)}
\label{sec:visualization choices}

In RQ2, we investigate how design differences across state-of-the art local explanation visualizations correlate with comprehension, bias perception, and trust. While we used 26~visualizations to build our taxonomy (see Figure \ref{fig:elements in tools}), we selected six visualizations from popular tools that span the vast majority of taxonomy design characteristics: SHAP waterfall plots~\cite{Lundberg17}, SHAP force plots~\cite{Lundberg18}, ELI5 tables~\cite{Korobov17}, LIME visualizations~\cite{Ribeiro16}, Ceteris-Paribus (CP) profile plots~\cite{Baniecki21}, and Anchors explanations~\cite{RibeiroTulio18}. We do so to avoid redundancy and diminishing returns, as many visualizations share overlapping characteristics

For each visualization, we conduct a survey assessing comprehension, bias perception, and trust for the same seven loan-recommendation scenarios. To ensure ecological validity, we did our best to keep the visualizations as similar as possible to those produced by each state-of-the-art tool. However, due to the multiple configuration parameters available, combined with heterogeneous approaches to computing and ordering feature importance~\cite{Krishna22}, we had to standardize non-visualization-related differences so that we can make conclusions regarding the effect of visual design. For example, we ensured that all visualizations had the same feature importance ordering and contribution values for a given scenario (using the SHAP order as our default). We also made minor standardization changes to facilitate understanding of the impact of visual design, rather than textual differences such as terminology (e.g., ``base value'' vs.\ ``bias value''), or capitalization. All of our standardizations are in our supplementary materials~\cite{supp}. 

Figure~\ref{fig:vis-combined} shows an example of each explanation visualization, as included in our surveys. We briefly describe each, highlighting key taxonomy characteristics motivating inclusion in our user study:

\begin{figure*}[t]
    \centering
    \begin{subfigure}{0.3\linewidth}
        \centering
        \includegraphics[width=\linewidth]{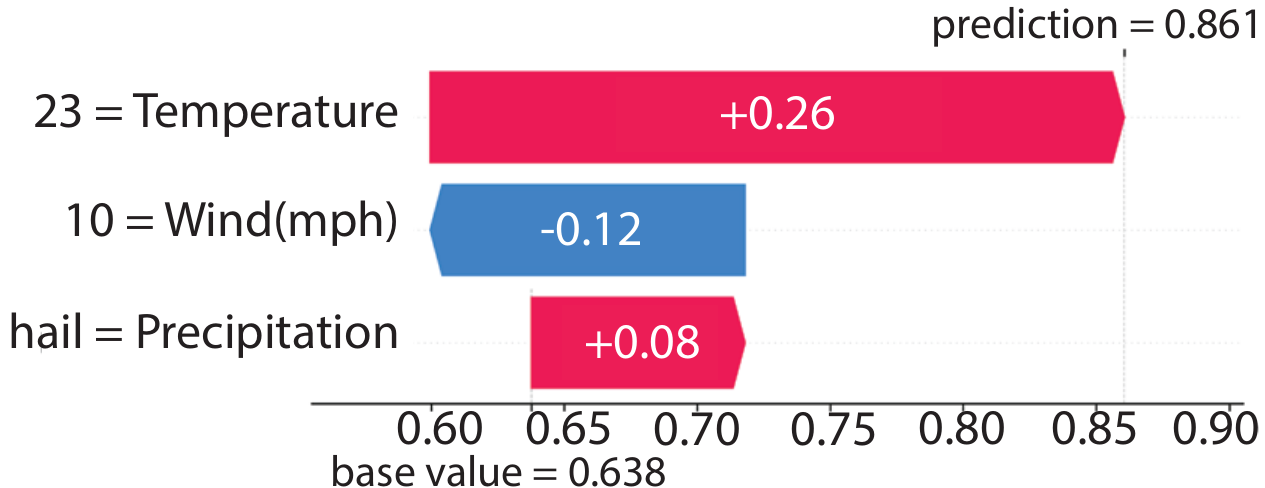}
        \caption{A SHAP waterfall plot~\cite{Lundberg17}.}
        \label{fig:shap-fig}
    \end{subfigure}
    \vrule width 0.5pt
    %\hfill
    \begin{subfigure}{0.3\linewidth}
        \centering
        \includegraphics[width=\linewidth]{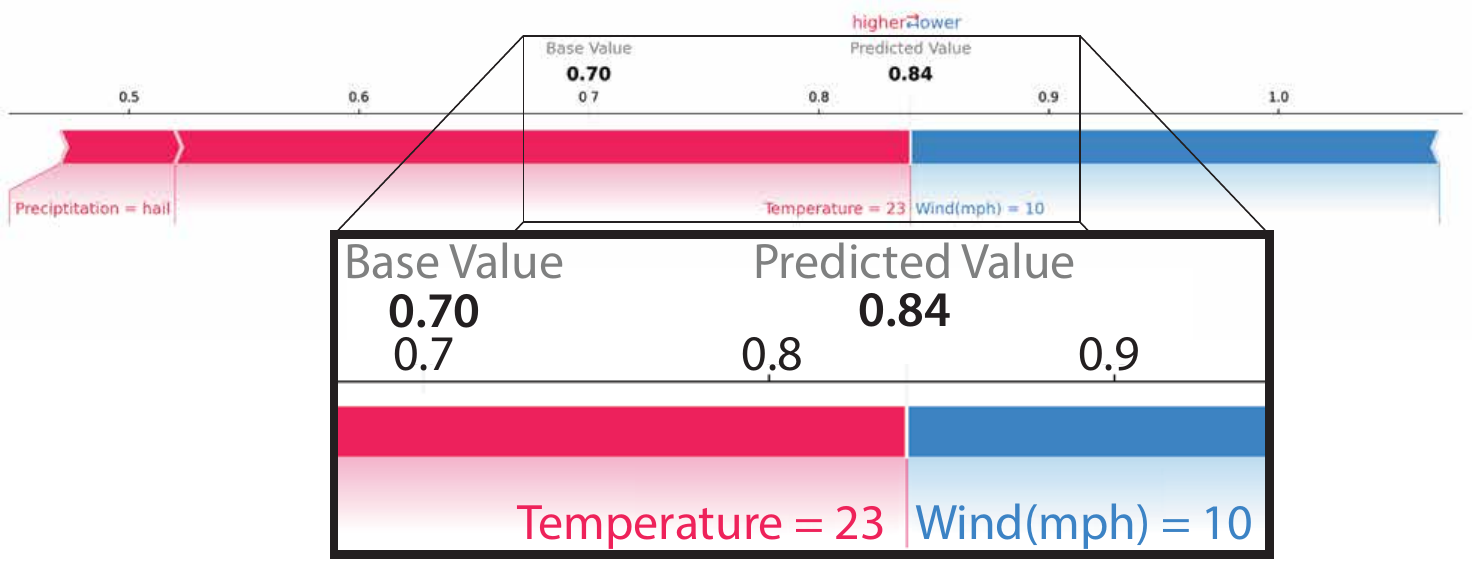}
        \caption{A SHAP force plot~\cite{Lundberg18}.}
        \label{fig:shap-force-fig}
    \end{subfigure}
    \vrule width 0.5pt
    \begin{subfigure}{0.3\linewidth}
        \centering
        \includegraphics[width=\linewidth]{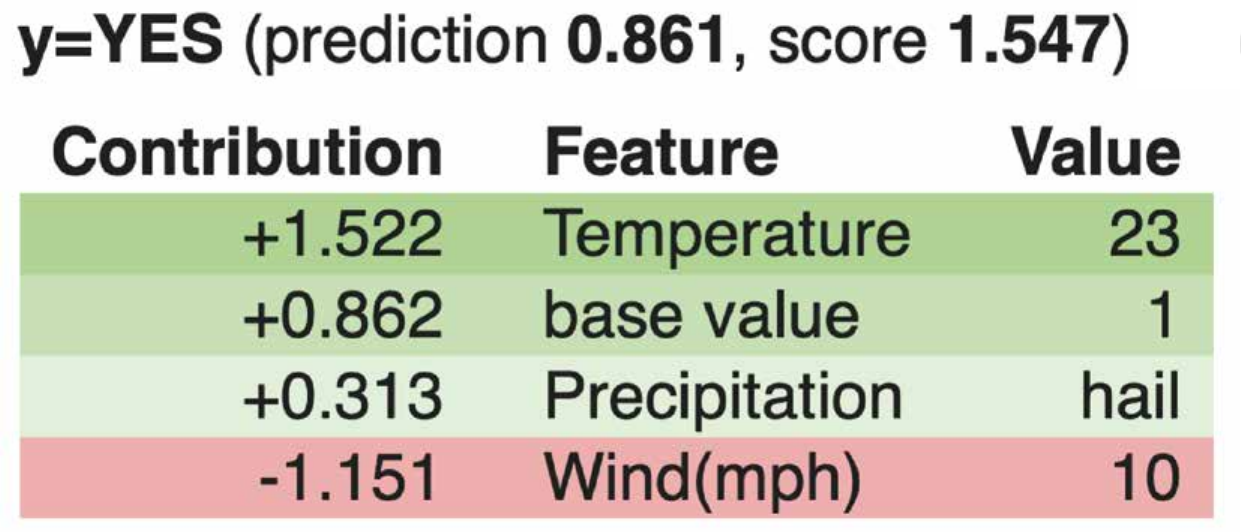}
        \caption{An ELI5 table~\cite{Korobov17}.}
        \label{fig:eli5-fig}
    \end{subfigure}
    \hrule
    \begin{subfigure}{0.3\linewidth}
        \centering
        \includegraphics[width=\linewidth]{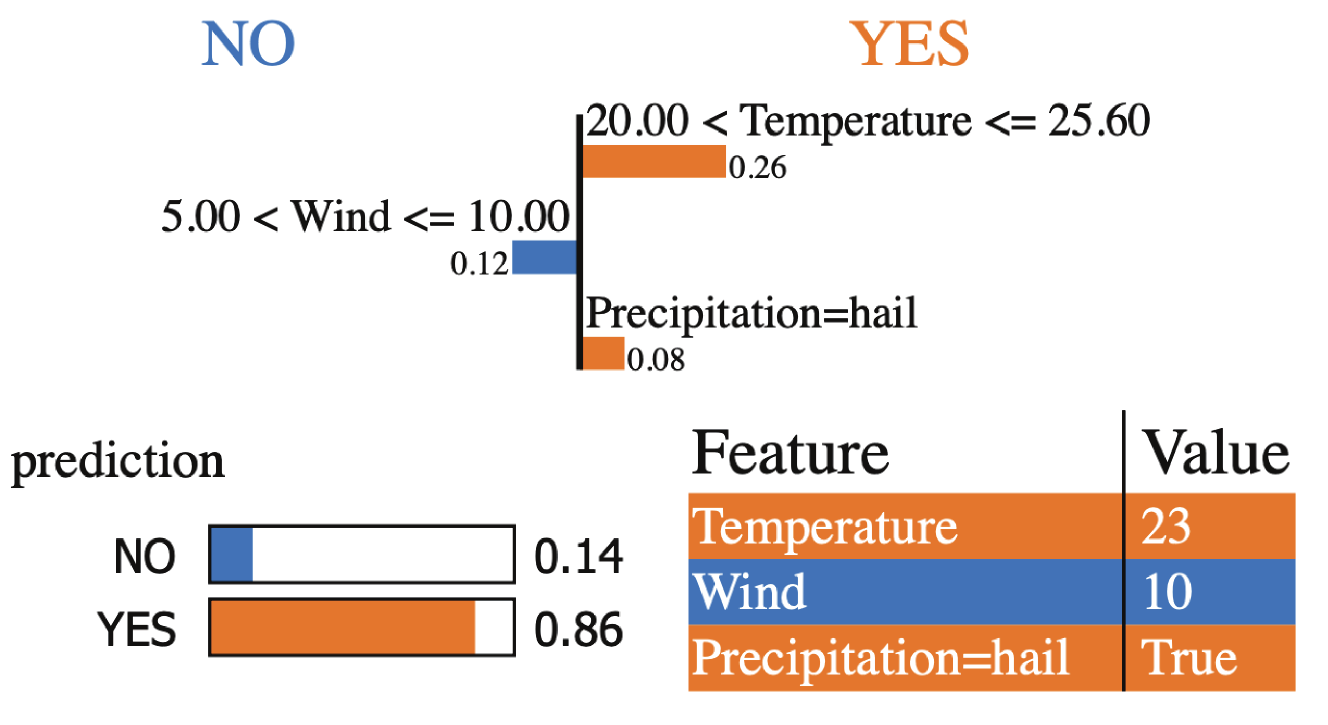}
        \caption{A LIME visualization~\cite{Ribeiro16}.}
        \label{fig:lime-fig}
    \end{subfigure}
    \vrule width 0.5pt
    %\hfill
    \begin{subfigure}{0.3\linewidth}
        \centering
        \includegraphics[width=\linewidth]{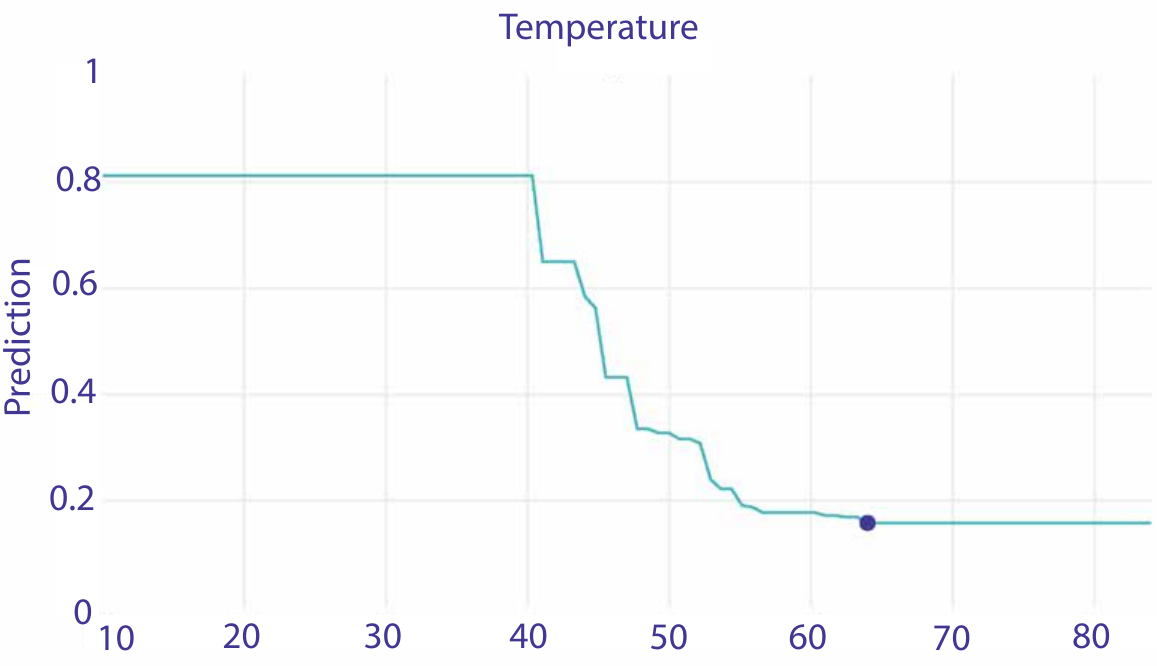}
        \caption{A Ceteris-Paribus (CP) profile plot~\cite{Baniecki21}.}
        \label{fig:cp-fig}
    \end{subfigure}
    \vrule width 0.5pt
    \begin{subfigure}{0.3\linewidth}
        \centering
        \includegraphics[width=\linewidth]{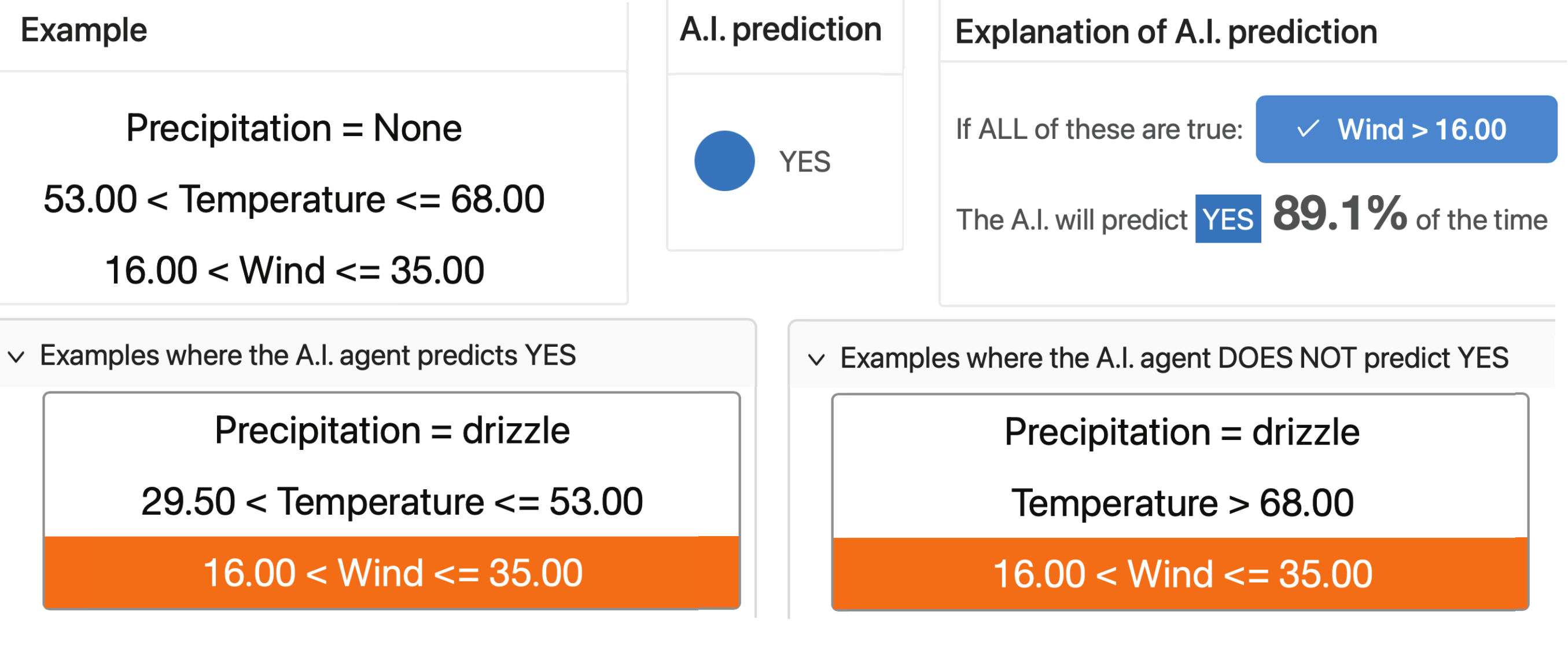}
        \caption{An anchors explanation~\cite{RibeiroTulio18}.}
        \label{fig:anchors-fig}
    \end{subfigure}
    \caption{The six state-of-the-art explainability visualizations in RQ2's user study: (a)~SHAP waterfall, (b)~SHAP force, (c)~ELI5, (d)~LIME, (e)~CP, and (f)~Anchors. We conducted six surveys on the same seven loan-recommendation scenarios, one with each visualization. Here, each explains the same input for a ML model that predicts if one should wear a coat based on the weather. We used this model in a tutorial at the start of our surveys. }
    \label{fig:vis-combined}
\end{figure*}

\emph{SHAP waterfall plots, Figure~\ref{fig:shap-fig}:} SHAP value attribution shows each feature's conditional contribution to the model's output, summing to the final predicted probability. SHAP is based off of shapley values in game theory, which are the solution to the equation $p(\text{output}) = b + \phi_1(\text{feature}_1) + \phi_2(\text{feature}_2) + \phi_3(\text{feature}_3) + \cdots$, where each $\phi$ is a shapley value.
 In the waterfall plot~\cite{Lundberg17}, colored arrow bars indicate positive (red) and negative (blue) feature contributions, stacked bottom-to-top from least to most significant. Starting at a base value, contributions shift the cumulative sum right (positive) or left (negative), with the top bar showing the final predicted probability.
 SHAP waterfall is one of only two visualizations in our study where bar elements indicate classifier output. 

\emph{SHAP force plots, Figure~\ref{fig:shap-force-fig}:} These show the same information as waterfall plots, but as a single bar where red (positive) and blue (negative) sections represent feature contributions~\cite{Lundberg18}. Sections are ordered around the predicted value, with positive contributions left and negative right.  More significant features are closer to the center. 
    %and a scale supports visual estimation. 
    SHAP force is the only visualization in our study where horizontal position indicates feature impact magnitude.

\emph{ELI5, Figure~\ref{fig:eli5-fig}:} A Python package for explaining classifier predictions, ELI5 shows feature contributions as a table with colored rows: large positive contributions at the top (deep green) and large negative ones at the bottom (deep red).  Each feature has a contribution score, computed in a model-specific manner. 
    For LightGBM, ELI5 traces ensemble decision tree paths to determine how each step impacts the prediction. Contributions sum to an output score, though this score does not directly represent the model's predicted probability~\cite{Korobov17}. 
    ELI5 is the only visualization in our study where vertical position indicates feature impact direction, and one of only two where classifier output is explicit or where color indicates feature impact magnitude. 
    
    %\item 
    \emph{LIME, Figure~\ref{fig:lime-fig}:} % (Local Interpretable Model-agnostic Explanations) 
    LIME explains model predictions by (1~generating ``near-by'' input-output pairs by input mutation, and (2)~fitting a white-box linear regression to estimate feature weights~\cite{Ribeiro16}. It visualizes feature weight magnitude and direction as a bidirectional horizontal bar chart, with features ordered by impact. Positive contributions are orange and point left; negative ones are blue and point right. 
    LIME also includes a bar chart with class prediction probabilities, and a table where row color and placement show feature significance. LIME is one of only two visualizations in our study where feature values are binned or bar elements indicate classifier output.

\emph{CP profiles, Figure~\ref{fig:cp-fig}:} In a separate plot for each feature, CP profiles show how changing a feature affects model output, while holding other features constant (line graphs for continuous features, bar charts for categorical ones)~\cite{Biecek21}. While not explicit, average output change, measured via CP oscillation, reflects feature impact~\cite{Biecek21}. Our tutorial explained how to estimate this, with larger oscillations indicating stronger impact.
 We use the Python Dalex implementation~\cite{Baniecki21}, where dark blue marks the current input-output and teal lines or bars show how output varies by input. 
CP is one of only two visualizations in our study where feature impact is inferred, and the only one to not use color or text to indicate feature impact direction. 

\emph{Anchors, Figure~\ref{fig:anchors-fig}:} Anchors explain predictions by identifying feature constraints that, when satisfied, cause the model to consistently output the same prediction~\cite{RibeiroTulio18}. Each anchor is a rule specifying the constrained features and its associated precision. Explanations are primarily textual, with color (orange or blue) highlighting constrained features. It also includes example inputs both satisfying or violating the anchor, showing the effect on model output. 
Anchors is one of only two visualizations in our study where feature values are binned, and the only one where color indicates classifier outcome. 

\subsubsection{Adjusted Explainability Visualizations (for RQ3)}
\label{sec:RQ3Methods}

While our analysis for RQ2 can establish correlations between comprehension, bias perception, and trust that are grounded in taxonomy design characteristics, it cannot establish causality. Establishing causality is important to assess if findings are likely to generalize beyond specific existing explanation visualizations, thus informing the design of next-generation visualization designs.   

In RQ3, we consider three causal experiments, one regarding the importance of explicit values, one assessing the impact of model fairness, and one on the relationship between bias perception and trust.  We describe each experiment in detail in Section~\ref{sec:RQ3}. However, at a high level, each involves two visualization surveys that are identical, other than a single visualization characteristic or model property, admitting causal comparison. Figure~\ref{fig:visRQ3-combined} shows examples of each pair of visualizations. While six surveys are analyzed in RQ3, we only run five additional surveys; since one condition for our explicitness experiment is a non-modified CP profile, we reuse its survey results from RQ2.

\subsection{Recruitment and Population Contextualization}
\label{sec:Participants}

We recruited 825 participants from the crowdsourcing platform Prolific.com~\cite{Palan18} who met our inclusion criteria (fluent in English, at least 18), 75 per survey. After additional quality filtering beyond Prolific's guarantees (e.g., attention checks, skipping questions), we had 818 valid participants (age 18--77); 438 for RQ2, and 380 for RQ3. 

Our sample size was guided by power analyses on preliminary data ($n$=50). Using balanced one-way analyses (power 0.9, effect size from pilot data, assumed normality), we estimated that 35--68 participants per visualization would be sufficient to detect effects for individual survey questions. While our final analysis used composite scores instead, this approach gave us confidence that we would have enough statistical power to convincingly answer our research questions.

We used Prolific's interface to recruit equal numbers of participants identifying as men and women, as prior work suggests that gender may influence bias perception and trust~\cite{Gaba23vis}. 397 identified as cis or trans women, 395 as cis or trans men, 19 as non-binary, 2 as an unlisted gender, and one as unsure/questioning. 
Participants varied in education, income, and ethnicity. Regarding ML familiarity, participants were primarily non-experts: 219 had none, 389 were beginners, 189 had intermediate knowledge, and 22 were experts.
Participants were compensated \$12.00 an hour, consistent with Prolific recommendations. 

\section{RQ2: Does visualization design correlate with comprehension, bias perception, and trust?}
\label{sec:RQ2}

Having defined key characteristics of local explanation visualizations (see Section~\ref{sec:Taxonomy}), we now examine if these characteristics correlate with model comprehension, bias perception, and trust. We conduct user studies with six state-of-the-art explainability visualizations, carefully selected to maximize coverage of our taxonomy: SHAP waterfall plots, SHAP force plots, ELI5, LIME, CP profiles, and Anchors (see Section~\ref{sec:Methodology} for study details and Section~\ref{sec:visualization choices} for visualization descriptions). Prior work has shown that model descriptions and transparency can affect perceived model fairness and trustworthiness~\cite{Wang20, Gaba23vis, xiong2019examining}. 

In this section, we analyze how each visualization relates to comprehension, perceived bias, and trust. To provide additional insight, we also consider higher-order connections between these properties. We conclude with a qualitative analysis of free-response answers. Overall, we find that visualization design significantly affects viewer comprehension, bias perception, and trust. Notably, higher comprehension is associated with lower trust. On investigation, we find that this relationship is mediated by bias perception---when the underlying model is biased, people are less likely to trust it.

\textbf{\textit{Comprehension} and Visualization Design.}
To assess how design characteristics influence comprehension, we fit a linear model using the comprehension score defined in Section~\ref{sec:Measures}. 
A one-way ANOVA test indicates that visualization type is a significant predictor of comprehension ($p < 0.001$).
Participants achieved the highest comprehension scores with LIME, and the lowest with Anchors (emmeans: LIME $= 41.2$, SHAP waterfall $= 39.7$,  SHAP Force $= 38.7$, ELI5 $=36.3$, CP $= 23.9$, Anchors $=23.0$). Visualizations that explicitly showed the magnitude and direction of feature impacts had higher comprehension (vs.\ those where they had to be inferred, $38.97$ vs.\ $23.46$). This  has direct implications for visualization design (see Section~\ref{sec:ImplicationsandFW}), and we test this finding causally in RQ3 (see Section~\ref{sec:followupone}). 

Breaking down the comprehension score (Figure~\ref{fig:measures}), we find that visualization is a significant predictor for each component (C1: $p=0.002$, C2: $p<0.001$, C3: $p<0.001$). Participants with LIME visualizations were significantly more likely to correctly assess the loan decision (emmean=$7.73$), and determine feature impact direction (emmean=$27.8$) or magnitude (emmean=$5.70$). 
This indicates that LIME visualizations characteristics facilitate comprehension. %In qualitative responses, 
Participants highlighted LIME visual elements as helpful, including its bar chart with classifier output probabilities (taxonomy dimension D10) and feature impact table (D6, D11, D12, D13, and D16). Our follow-up causality experiments (Section~\ref{sec:followuptwo}) include these elements.

Finally, we examine if participants are aware of their own comprehension level. A linear model shows that perceived comprehension predicts objective comprehension ($p < 0.001$). However, this effect is small (Pearson's $r=0.33$). This underscores the importance of visualizations that facilitate actual, rather than perceived, comprehension; model behavior can be incredibly complex, and non-experts may struggle to recognize gaps in their understanding. 

\textbf{\textit{Bias Perception} and Visualization Design.} 
Using a linear model, we find that visualization type significantly predicts perceived bias ($p=0.004$, emmeans: LIME $= 14.2$, SHAP Waterfall $= 12.6$, SHAP Force $= 15.0$, ELI5 $= 13.9$, CP $= 10.5$, anchors $=11.9$). Participants were least likely to perceive bias with CP and Anchors visualizations that have implicit feature impacts, and include alternative outputs (D16, D17, and D22). This indicates that bias perception may be facilitated by simpler visualizations with explicit feature impacts. We use this finding to design a follow-up experiment investigating the impact of manipulating bias perception on trust (see Section~\ref{sec:followupthree}). 

\begin{figure}[t]
    \centering
    \includegraphics[trim={0 0.15cm 0 0.5cm},clip,width=0.75\columnwidth]{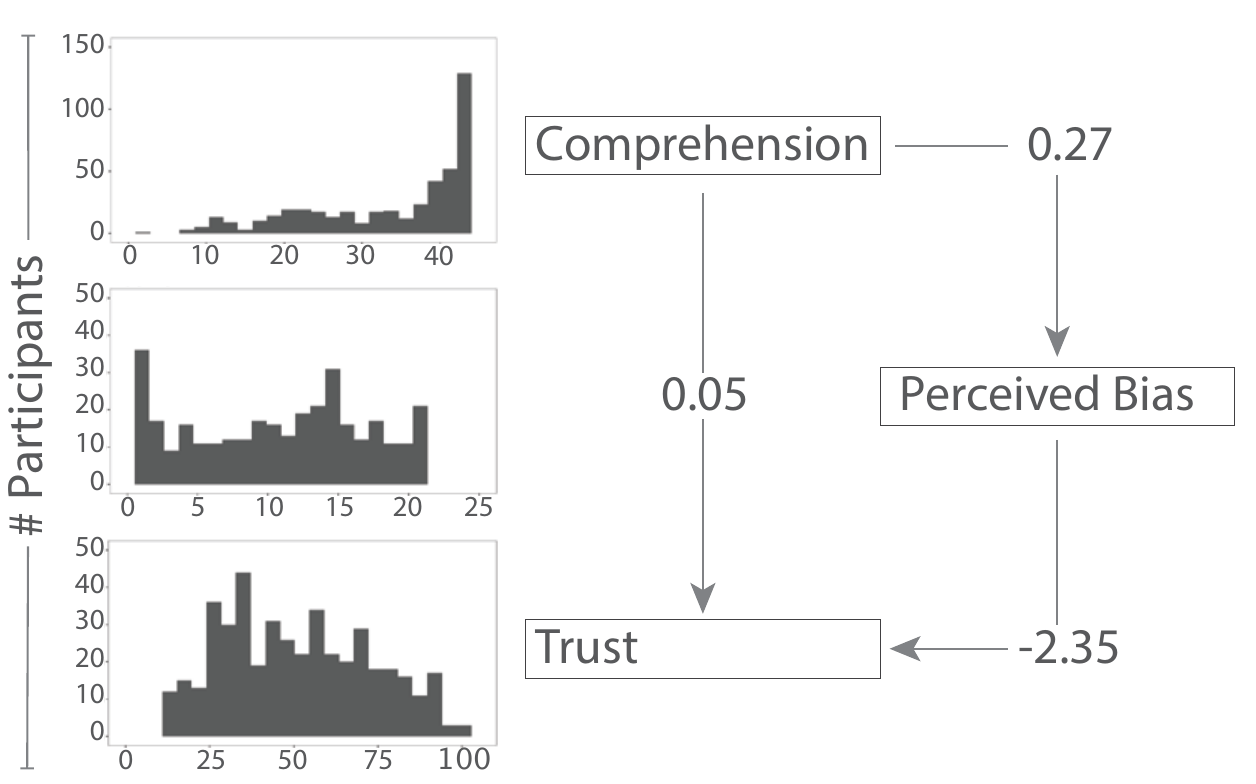}
    \caption{The mediation effects between comprehension, perceived bias, and trust. 
    While comprehension is negatively correlated with trust, this graph shows that this correlation is heavily mediated by perceived bias. The (estimated) direct effect of comprehension on perceived is significantly positive~($0.27$), and the direct effect of perceived bias on trust is significantly negative~($-2.35$), but the estimated direct effect of comprehension on trust is actually slightly positive at $0.05$. Furthermore, this last relationship is not statistically significant.}
    \label{fig:medgraph}
\end{figure}

\textbf{\textit{Model Trust} and Visualization Design.}
Visualization design also significantly predicts trust  ($p=0.03$). 
On average, participants trusted CP visualizations most and SHAP force plots least (emmeans: LIME $= 50.2$, SHAP Waterfall $= 52.0$, SHAP Force $= 45.8$, ELI5 $= 51.4$, CP $= 58.3$, Anchors$=53.9$). The high trust in CP may relate to its implicit design and alternative outputs, while
SHAP Force plots' unique horizontal bar elements (D10) may negatively impact trust.

\textbf{Higher-Order Connections: Comprehension, Bias Perception, and Trust.} To provide additional insight, we also consider higher-order connections between comprehension, bias perception, and trust. We observe a small but significant negative correlation between comprehension and trust (Pearson's $r = -0.28$, $p < 0.001$); participants who understood the model better tended to trust it less. 

Further analysis reveals this relationship is potentially mediated by bias perception.  
When plotting the relationship between comprehension, bias, and trust (see Figure~\ref{fig:teaser}), we notice a direct inverse relationship between the level of bias perception and trust across tested visualizations. That is, visualizations with higher bias perception also have lower trust. 
A linear model with trust as the dependent variable and both comprehension and bias perception as the predictors shows that  
\textit{bias perception alone predicts trust} (Sum sq $= 122260$, $p < 0.001$), while comprehension is not significant. 
This is interesting given our finding that comprehension score is negatively correlated with trust. 

To better understand this result, we fit a second model with bias perception as the dependent variable, and comprehension as the predictor. 
In contrast, this model shows comprehension is a significant predictor of bias perception (Sum sq $= 3577.3$, $p < 0.001$),
suggesting that there is potentially a heavy mediation effect of bias perception on the correlation between comprehension score and trust. 

To confirm this interpretation, we fit a mediation model with trust score as the dependent variable, the comprehension score as the predictor, and the bias perception score as the mediator. As Figure~\ref{fig:medgraph} shows, this analysis reveals that (1)~comprehension positively predicts bias perception (estimated coefficient $b=0.27$, $p<0.001$), meaning that increasing comprehension by one unit increases perceived bias by 0.27 units on average, (2)~bias perception negatively predicts trust ($b = -2.35$, $p < 0.001$), (3)~the direct effect of comprehension on trust is small and non-significant ($b=0.05$, $p=0.49$), and (4)~the indirect effect of comprehension on trust through bias perception is $b=-0.63$.

This suggests that the negative link between comprehension and trust is heavily mediated by bias perception: higher comprehension makes bias more apparent, reducing trust. Importantly, this also implies that bias perception—and therefore trust—can potentially be modulated independently of comprehension, motivating the design of visualizations that support both comprehension and bias awareness (see Section~\ref{sec:ImplicationsandFW}).

\label{sec:behavioral alignment}

\begin{figure}[t]
    \centering
    \includegraphics[trim={0 .55cm 0 .65cm},clip,width=0.90\columnwidth]{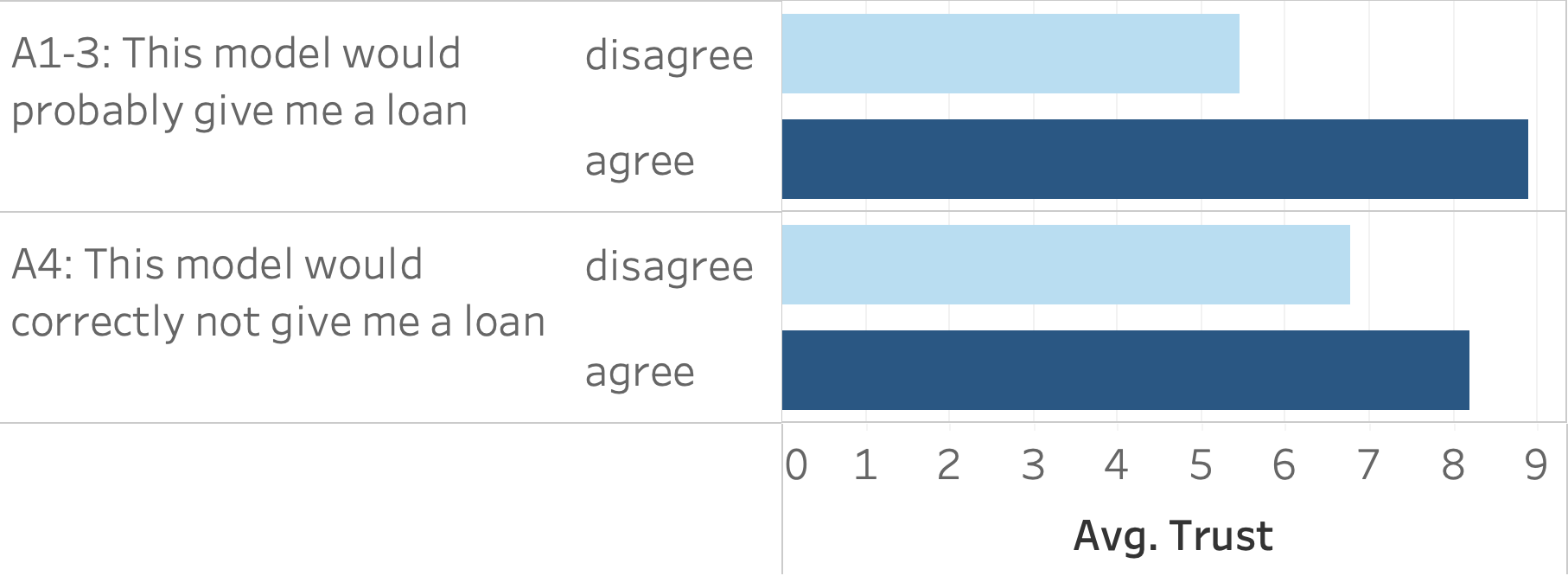}
    \caption{A bar graph showing the difference in average trust between instances where participants felt the model would give them a loan, and instances where they felt it would not. Average trust here is by instance rather than aggregated. A second set of bars shows the difference in average trust between instances where participants agreed with the statement that the model rightfully would not give them a loan, and instances where they disagreed. Corresponding statements can be found in Figure~\ref{fig:measures}.}
    \label{fig:visandtrust}
\end{figure}

\textbf{Behavioral Alignment.} %Previous work has found that 
People are more likely to trust models that benefit them, even if those models are biased~\cite{Gaba23vis}.
We investigate if participants' relationship to the model (e.g., if they expect it to give them a loan) affected trust. Participants believed the model would give them a loan in about half of all instances ($52.15\%$). In one third ($33.86\%$), participants believed the model would not give them a loan but that this would be the correct choice. Comparing trust scores, we find that participants trusted the model significantly more when they believed it would give them a loan ($p<0.001$). However, Figure~\ref{fig:visandtrust} shows that when participants believed they would not receive a loan but agreed with this outcome, there was again a significant (though smaller, $1.54$ vs.\ $3.72$) increase in average trust ($p<0.001$). We define a person holding either of these beliefs about the model to be in behavioral alignment. We find that behavioral alignment has a large  positive effect on trust (Cohen's $d = 1.26$) and a large  negative effect on bias perception (Cohen's $d = -0.98$). These findings imply that favorable or agreeable outcomes can reduce users' perception of bias and increase trust, regardless of the model's actual fairness.

\begin{figure}[t]
    \centering
    \includegraphics[width=0.8\columnwidth]{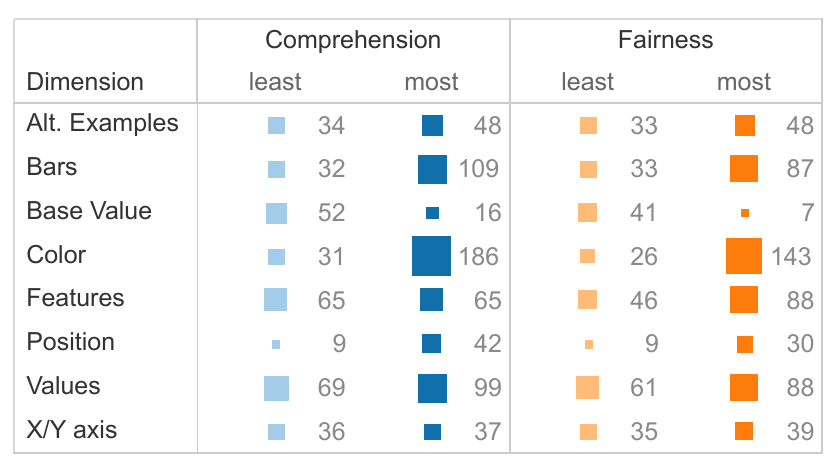}
    \caption{A bubble chart showing the number of participants mentioning taxonomy features when asked free-form questions regarding the characteristics they found most and least helpful when answering comprehension and fairness questions. These features are grouped by meta-dimension type. Explicitness is excluded as it is not directly mentioned.}
    \label{fig:elementimpvis}
\end{figure}

\textbf{Qualitative Analysis.} Finally, we asked participants free response questions regarding which visualization characteristics they found most and least useful, to further inform the design of next-generation explanation visualizations. Via manual analysis, we aggregate responses across taxonomy dimensions. Figure~\ref{fig:elementimpvis} summarizes the number of participants that found each dimension to be most and least helpful when assessing model comprehension or fairness. 
While participants perceived almost all dimensions as more useful than not, they found color-coding impact direction and magnitude to be particularly useful. However, anecdotally, for both SHAP Plots, participants found red for positive contributions and blue for negative contributions to be counterintuitive. Furthermore, both bars and explicit values were found to be useful. Conversely, numbered x/y axes were not found to be particularly useful in conveying information---participants preferred simpler visual cues like color or bar size. Very few participants found base values useful. 
These qualitative observations motivate the design of our second crowd-sourced controlled experiment in RQ3 (Section~\ref{sec:followuptwo}). 

\summarytcolorbox{RQ2 Summary: Visualization Design Correlations}{We find evidence suggesting that comprehension, trust, and bias perception are all affected by visualization design. For example, viewers trusted the underlying model less when given explanations with more explicit information. 
We further find a negative correlation between comprehension and trust, which is heavily mediated by the perception of bias; when people understand a model more, they trust it less, potentially due to increased bias perception.} 

\section{RQ3: Confirmation of Causality}
\label{sec:RQ3}

\if False % Zhannas
\begin{figure*}[t]
    \centering
    \begin{subfigure}{0.18\linewidth}
        \centering
        \includegraphics[width=\linewidth]{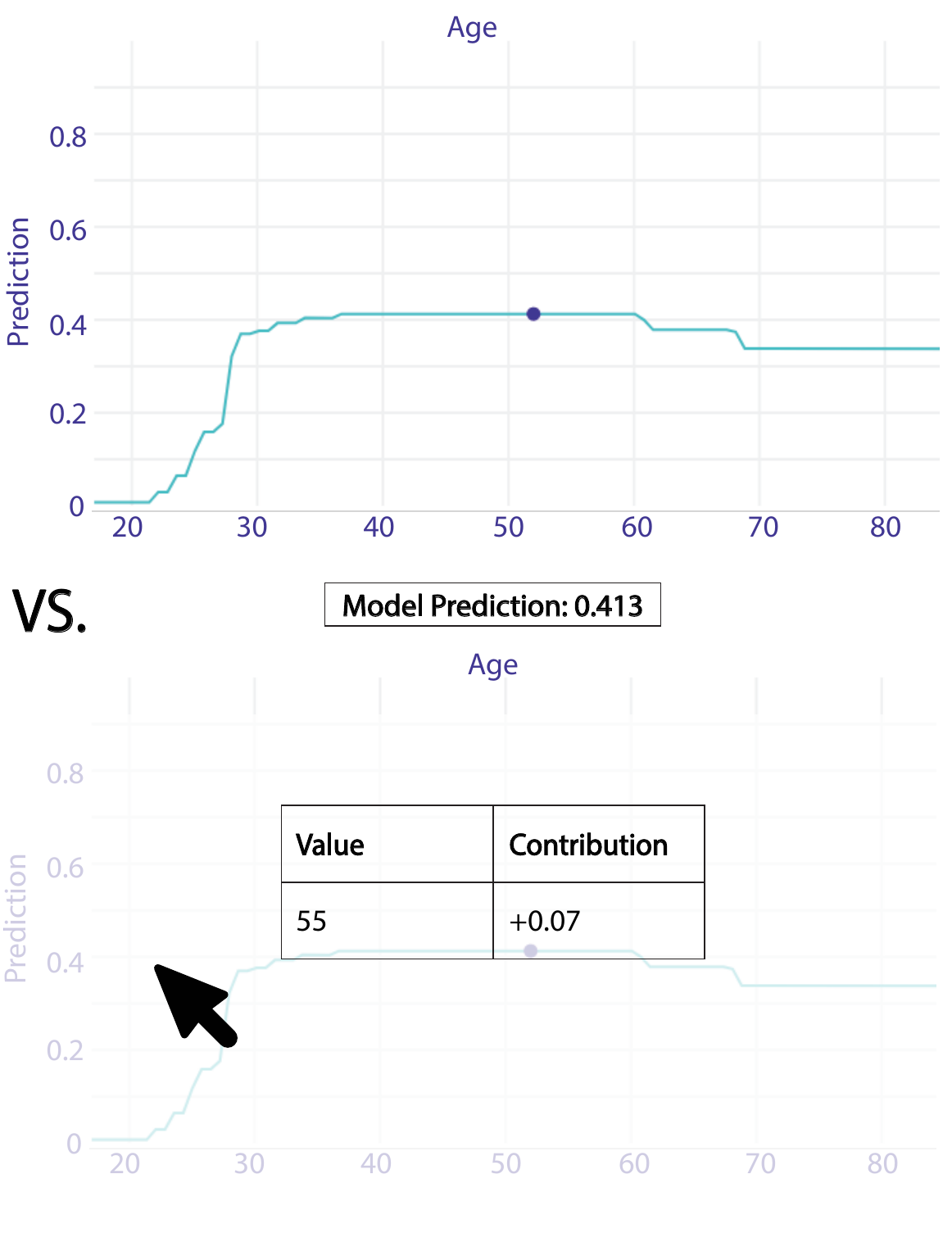}
        \caption{Follow-up experiment comparing implicit and explicit information presentation.}
        \label{fig:explicit-exp-fig}
    \end{subfigure}
    \hfill
    \begin{subfigure}{0.40\linewidth}
        \centering
        \includegraphics[width=\linewidth]{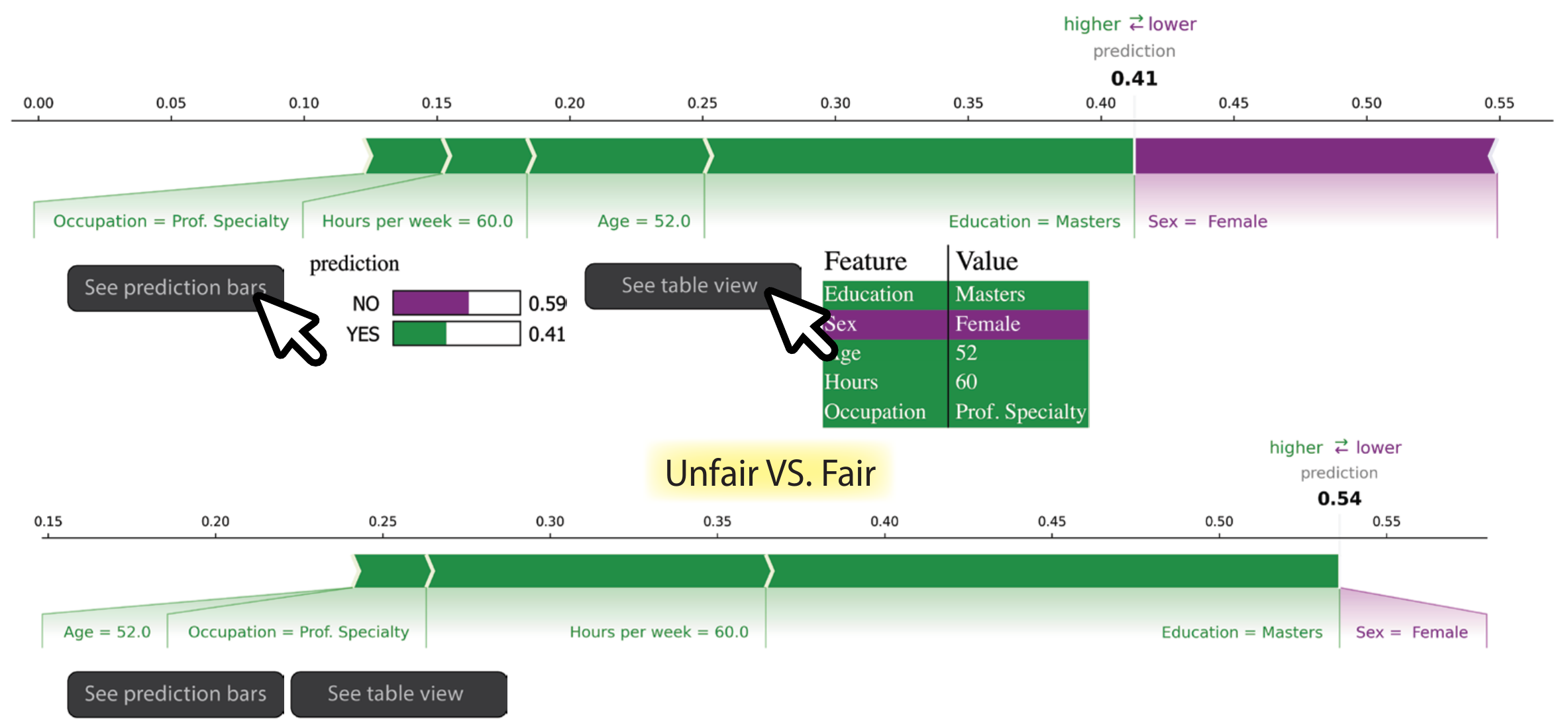}
        \caption{Follow-up experiment two comparing a fair and unfair underlying model.}
        \label{fig:fair-exp-fig}
    \end{subfigure}
    \begin{subfigure}{0.40\linewidth}
        \centering
        \includegraphics[width=\linewidth]{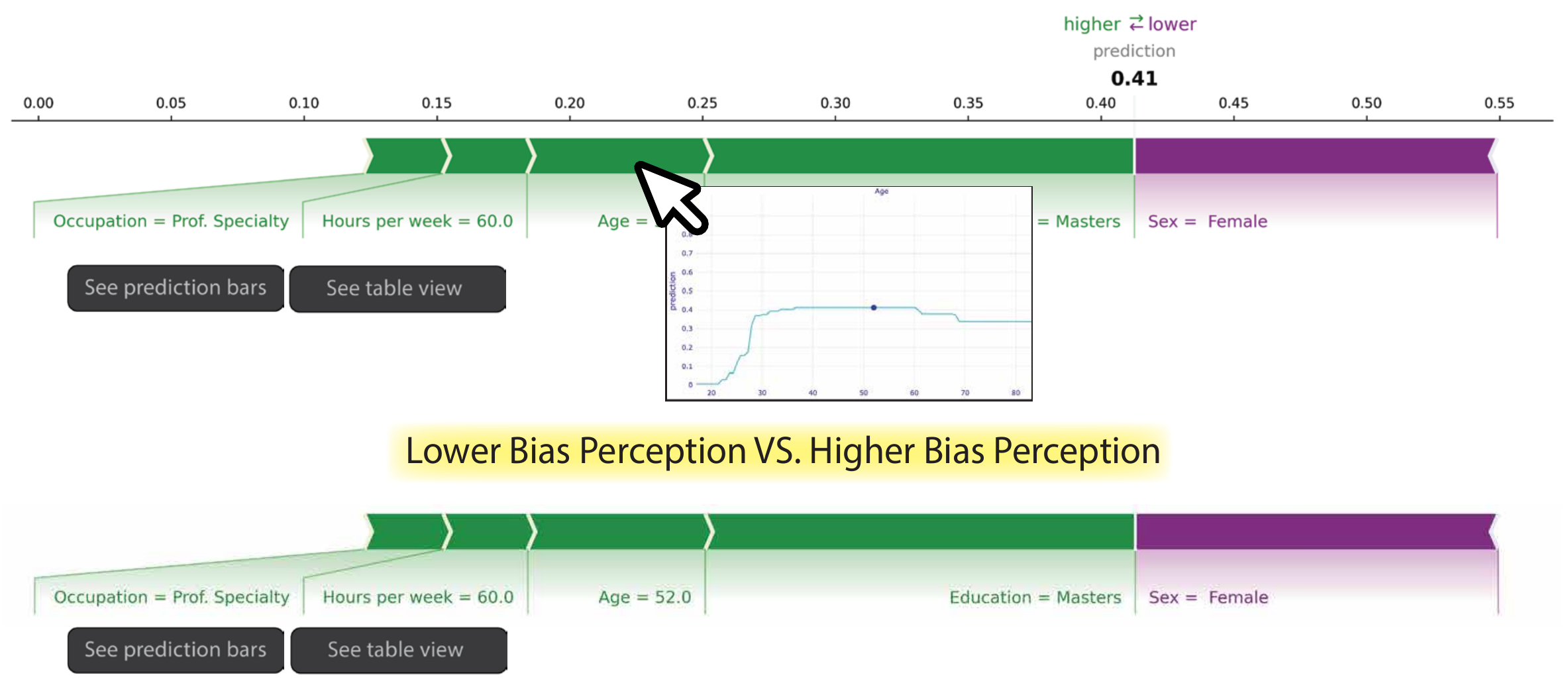}
        \caption{Follow-up experiment comparing visualizations with artificially lowered and increased bias perception.}
        \label{fig:bias-percep-exp-fig}
    \end{subfigure}
\end{figure*}

\fi

\begin{figure}[t]
    \centering
    \begin{subfigure}{\linewidth}
        \centering
        \includegraphics[width=0.9\linewidth]{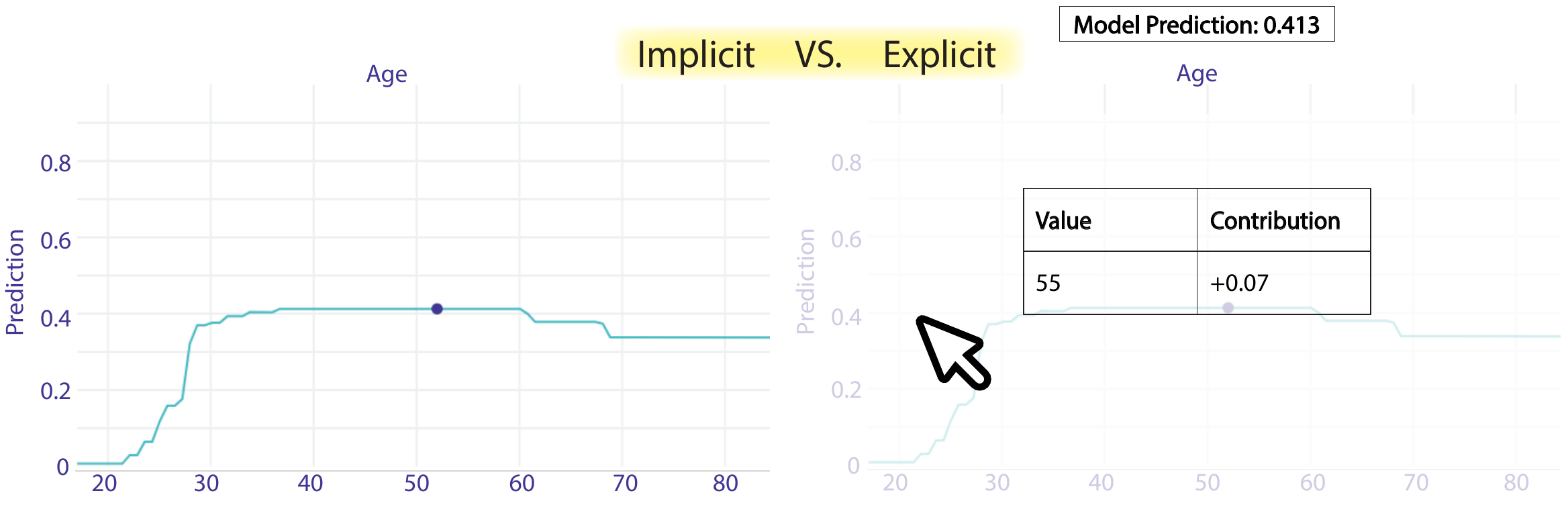}
        \caption{\emph{Experiment 1:} Compares implicit and explicit feature values. We compare responses for a standard CP plot (left) to a CP plot that we manually augmented to show explicit feature values when moused over (right).} 
        \label{fig:explicit-exp-fig}
    \end{subfigure}
        \hrule

    \begin{subfigure}{\linewidth}
        \centering
        \includegraphics[width=0.9\linewidth]{figs/fair_exp.pdf}
        \caption{\emph{Experiment 2:} Compares responses to an unfair (top) and fair (bottom) model, using a composite visualization designed for high comprehension. The top model uses sex as a significant feature for a loan-recommendation prediction, while the bottom one does not.}
        \label{fig:fair-exp-fig}
    \end{subfigure}
    \hrule

    \begin{subfigure}{\linewidth}
        \centering
        \includegraphics[width=0.9\linewidth]{figs/perceive_bias_exp.pdf}
        \caption{\emph{Experiment 3:} Compares responses to composite visualizations facilitating lower (top) and higher (bottom) bias perception. The top visualization adds CP plots on mouse-over of each feature to show alternative outputs and increase visualization complexity, while the bottom removes the x and y axes to increase simplicity.}
        \label{fig:bias-percep-exp-fig}
    \end{subfigure}
    \vspace{-5ex}
    \caption{Three crowd-sourced controlled experiments testing for causality in between comprehension, bias perceptions, and trust by manipulating comprehension~(a), underlying bias~(b), and perceived bias~(c). 
    \vspace{-3ex}}
    \label{fig:visRQ3-combined}
\end{figure}

In RQ2 (Section~\ref{sec:RQ2}), we observed a correlation between comprehension and trust, mediated by perceived model bias. We also observed that explicit visualization of feature impacts improves comprehension and bias perception. To test the causal nature of these relationships, we conduct three crowd-sourced controlled experiments using the 7 prediction instances from Section~\ref{sec:ML model and its instances}. We use our taxonomy's relevant visualization characteristics to create pairs of identical visualizations, except for the key attribute under test (see Figure~\ref{fig:visRQ3-combined} for experimental visualization pairs).
This approach allows us to establish causality and test if observed correlations generalize beyond specific tools. 

\vspace{2pt}
\noindent \emph{Experiment 1---Explicitness:} We investigate the explicit feature impact of magnitudes and direction on comprehension and bias perception. This experiment is motivated by our finding in RQ2 that visualizations with explicit information about feature impact and direction had higher comprehension scores. To compare explicit and implicit visualizations, we compare variations of the CP survey described in Section~\ref{sec:visualization choices} with and without mouse-over explicit values (See Figure~\ref{fig:explicit-exp-fig}). Given our observations in Section~\ref{sec:RQ2}, we hypothesize that adding explicitness will increase comprehension and bias perception, while reducing trust. 

\vspace{2pt}
\noindent \emph{Experiment 2---Fairness:}  We test if reducing model bias increases trust by lowering perceived bias, motivated by our findings in RQ2 that bias perception mediates the negative relationship between comprehension and trust. We hypothesize that for high-comprehension visualizations, reducing model bias will also reduce perceived bias and increase trust. We construct an adjustable composite visualization combining design characteristics associated with comprehension (e.g., LIME's probability bar and feature impact table) and bias perception (e.g., SHAP Force's bar chart), which users can explore through interactive mouse-overs. 

We then conduct two surveys with this composite visualization, one where participants see only the fair model, and one where participants see only the biased model (see Figure~\ref{fig:fair-exp-fig}). We chose to reduce bias for sex and age as these were the two protected characteristics from the existing set. We found that our original model gave a loan to 24\% of men in a test set, and only 1.4\% of women. Furthermore, our model gave 27\% of those over the age of 37 (median age) a loan vs.\ only 5.4\% of those under the age of 37.
To create our fair model, we used the Seldonian Algorithm~\cite{Thomas19science}, constraining model behavior to have demographic parity ($Pr(Y|\mathit{Group1}) - Pr(Y|\mathit{Group2})<\epsilon$) for both sex and age. Using the demographic parity metric allowed us to reduce bias in a way that vas visible in individual outputs and did not require a fair training dataset~\cite{Thomas19science}.
Using the Seldonian toolkit, we trained a random forest model that fits our fairness constraints while preserving accuracy (accuracy of $0.790$ vs.\ $0.814$ for the original model).

\vspace{2pt}
\noindent \emph{Experiment 3---Bias Perception:} Finally, we test if altering bias perception through visualization design can affect trust without changing model behavior. We adjust our composite visualization from Experiment 2 to decrease and increase bias perception, creating two new variations (see Figure~\ref{fig:bias-percep-exp-fig}). To decrease bias perception, we incorporate the characteristics identified in Section~\ref{sec:RQ2} as associated with the lowest bias perception by adding CP Plots on feature mouse-over. To increase bias perception, we maximize simplicity by removing characteristics qualitatively identified as unhelpful, such as the x and y axes. We hypothesize that reducing bias perception will increase trust, even if the model's biased behavior does not actually change.

\subsection{Experiment 1---Explicitness}
\label{sec:followupone}

We test whether making feature impact magnitudes and directions explicit affects comprehension, bias perception, and trust. Since our distributions for comprehension, perceived bias, and trust are not normal (see Figure~\ref{fig:medgraph}), we use Wilcoxon Rank sum tests for comparisons and Cohen's $d$ for effect size. 

We find that adding explicit indicators of model output, feature impact direction, and feature impact magnitude has a significant effect on all three measures ($p<0.001$). Specifically, explicitness has a positive medium effect on comprehension ($d=0.68$), a small positive effect on bias perception ($d=0.22$), and a small negative effect on trust ($d=-0.26$). This indicates that including explicit values increases comprehension and bias perception, and decreases trust. This result is exciting; not only does it demonstrate how our taxonomy can be used to learn deeper visualization insights, it also indicates that modulating taxonomized design characteristics can increase viewer comprehension of the underlying model, reveal model biases, and adjust viewer trust.

\subsection{Experiment 2---Fairness}
\label{sec:followuptwo}

We test if reducing model bias increases trust, in the context of a high-comprehension visualization. We create this visualization by combining elements of visualizations with high comprehension scores. We use a purple and green color scheme, modifying counterintuitive colors and ensuring they are color-blind friendly.
We see significant differences between participants who saw explanations of our unfair vs.\ fair models, across all measures including comprehension ($p=0.003$), trust ($p<0.001$), and bias perception ($p<0.001$). Using the fair model results in both a small negative effect on bias perception ($d=-0.33$), and also a small-medium positive effect on trust ($d=0.44$). While significant, the effect on comprehension is negligible ($d=-0.16$). 

These findings indicate that with high comprehension, decreasing true bias decreases bias perception, leading to an appropriate increase in trust. This result supports a direct causal relationship between bias perception and trust, and demonstrates that visualizations that facilitate comprehension also calibrate bias perception with true fairness, underscoring the importance of comprehension to visualization design.

\subsection{Experiment 3---Bias Perception}
\label{sec:followupthree}

Finally, we test if we can use visualization design to modify perceived bias and affect trust, even when the model's underlying behavior is unchanged. We observe significant differences in perceived bias ($p<0.001$) and trust ($p<0.001$) between participants who saw our two survey variations. By adjusting our composite visualization to include taxonomy characteristics correlated with decreased bias, we were able cause both a small decrease in bias perception ($d=-0.27$) and also a small increase in trust ($d= 0.20$). There was also a significant change in comprehension  ($p<0.01$), but the effect size was tiny ($d= -0.07$). 

These findings indicate that even in the case where changes in comprehension are negligible, visualizations designed to obscure model bias will increase viewer trust in that model. These observations support our hypothesis of a direct causal relationship between bias perception and trust, which can be manipulated through visualization design.

\summarytcolorbox{RQ3 Summary: Causal relationships}{\looseness-1 We find that increasing comprehension of a biased visualization leads to increased bias perception and decreased trust, while high comprehension of a fair visualization results in decreased bias perception and increased trust. However, even when underlying bias does not change, artificially decreasing bias perception can increase trust. These findings support causality for our observed results from RQ2.} 

\section{Limitations}
\label{sec:Limitations}

We operationalize our three major measures as described in Section~\ref{sec:Measures} based on existing work and the characteristics present in the chosen visualization set. 
However, there are other methods of measuring comprehension and trust, in particular. 
Possible comprehension questions will always depend in part on the information intended to be conveyed by the visualization designers. 
For CP, for instance, feature importance is not an inherent part of designer intent~\cite{Biecek21}. 
Therefore, there may be additional questions that better capture comprehension of this visualization that we were unable to ask because the same information was not present in LIME, SHAP, or ELI5. 
Trust in a model can be operationalized differently as well---for instance, through decision questions~\cite{Hoque22} or trust games~\cite{Gaba23vis}. 
These measures may more accurately capture trust.  

Our model included 6 input features to limit respondent fatigue, which could reduce participant engagement~\cite{Jeong23}. However, real-world classifier applications likely require larger feature input sets and can potentially involve numerous and overlapping biases. Furthermore, our primary model bias was gender-based, and gender-based AI bias is a well-documented social issue~\cite{Jerlyn25}, so it is possible that participants were primed to perceive the model as biased upon seeing gender as a feature. Scaling our experiments to more complex models or less widely recognized biases may require alternative survey methodology.

The explanation types of local explainability visualizations vary (see dimensions D24--D27 in Figure~\ref{fig:elements in tools}). For example, SHAP, LIME, and ELI5 present additive feature contribution values, while Anchors presents if-then rules, and CP profiles present outcome sensitivity to individual input feature changes. The type of information presented may impact both comprehension and perception of bias, and offering multiple explanation types in combination may result in a deeper understanding of model functionality. Further research is necessary to better understand how different explanation types both individually and in combination can impact viewers of explainability visualizations.

We neither varied the model's level of bias, beyond creating one fair and one unfair version, nor the model's comprehensibility. Future work investigating more granular bias variations may provide more nuanced insights into the relationship between comprehension and trust.

\section{Discussion and Design Implications}
\label{sec:ImplicationsandFW}

We find that the types of visualization characteristics used to impart local explainability information affect a user's comprehension of the underlying model, their perception of bias in that model, and their trust of that model. Explicitly indicating feature importance information, via color or printed values, can increase both comprehension and perception of bias. Qualitative results show that people prefer certain characteristics over others when trying to comprehend model behavior and perceive model bias, and that people are more likely to trust a model they feel would benefit them. Quantitatively, we find a negative correlation between participants' comprehension of and trust in ML models, strongly mediated by the perception of bias. In other words, when dealing with biased models, better visualizations lead people to more greatly perceive bias, reducing trust. However, we also found that certain design decisions can alter bias perception without affecting comprehension. Anecdotally, we notice that visualizations with lower complexity and more explicitness correlate with higher bias perception and lower trust, and this correlation should be explored in future work. Furthermore, the very presence of input features like sex and age resulted in some participants seeing the model as discriminatory, even in the case of the fair model. Our findings indicate that explainability visualizations, when carefully designed, can be a useful tool in revealing ML model behavior and bias to a variety of users, including non-experts. We caution explainability designers to consider the clarity and intuitiveness of their designs to variable user populations, as well as the potential for those designs to obscure problematic model behavior. We further suggest that AI developers looking to use explainability to debug bias in their models consider how different presentation methods may affect their understanding of their own models' functionality.

\looseness-1
Anecdotally, all 19 non-binary participants across survey variations found both fair and biased models to be discriminatory. Marginalized communities are more sensitive to discrimination~\cite{Lee21}, but there is a dearth of research into non-binary individuals' perception of ML bias. Our use of a dataset and model with only two genders may have made non-binary participants feel excluded, possibly affecting their perception of the model. Unfortunately, real-world datasets with data for non-binary individuals are not readily available.  Future work should look explicitly at non-binary individuals' perception of bias in technology.

\section{Contributions and Future Work}
\label{sec:Contributions}

ML-powered systems are increasingly common, but they are often biased. Explanation visualizations may help non-ML-expert stakeholders understand and assess model outputs, but there is a limited understanding of how visualization design can systematically impact user perception. 
We take steps towards improving our understanding of how explainability visualization design impacts ML model comprehension, bias perception, and trust. 
First, we survey local explainability visualizations, to create a taxonomy of visualization design characteristics.
We then conduct a series of user studies, identifying correlational and causal relationships regarding how these characteristics facilitate comprehension, bias perception, and trust.
Our results provide insights for next-generation visualization tools that can better empower stakeholders to make well-informed, responsible decisions about ML applications. Our work forms an important step towards understanding how people's bias perception of ML outputs affects their trust, and how visualization techniques can help improve effectively communicating important aspects of ML models to non-expert, everyday users.

\acknowledgments{%
This work is supported by the National Science Foundation
under grant no. CCF-2210243.}
\bibliographystyle{abbrv-doi-hyperref-narrow}

\bibliography{softeng,laser,vis}

\end{document}